%% file: paper.tex
\documentstyle[aps,prb,epsf,amsmath]{revtex}

\begin{document}
\draft

\title{Random Magnetic Interactions and Spin Glass Order Competing with
Superconductivity: Interference of the Quantum Parisi Phase}
\author{H. Feldmann and R. Oppermann}
\address{Institut f. Theoret. Physik, Univ. W\"urzburg D--97074
W\"urzburg, {\it Federal Republic of Germany}}
\date{\today}

\maketitle


\input{abs}         
\input{intro}       
\input{results}     
\input{model}       
\input{scsgbasics}  
\input{sg1rsb}      
\input{sc1}         
\input{scM6}        
\input{scsg0rsb}    
\input{scsg1rsb}    
\input{pam}         

\bibliographystyle{prsty}
\bibliography{oppgroup}

\end{document}

%% file: abs.tex
\begin{abstract}
We analyse the competition between spin glass (SG) order and local pairing
superconductivity (SC) in the fermionic Ising spin glass with frustrated
fermionic spin interaction and nonrandom attractive interaction.
The phase diagram is presented for all temperatures $T$ and
chemical potentials $\mu$. SC--SG transitions are derived for
the relevant ratios between attractive and frustrated--magnetic
interaction. 
Characteristic features of pairbreaking caused by random magnetic
interaction and/or by spin glass proximity are found. 
The existence of low--energy excitations,
arising from replica permutation symmetry breaking (RPSB) in the Quantum
Parisi Phase, is shown to be relevant for the SC--SG phase boundary. 
Complete 1-step RPSB-calculations for the SG--phase are presented together
with a few results for $\infty$--step breaking.
Suppression of reentrant SG - SC - SG transitions due to RPSB is found and
discussed in context of ferromagnet - SG boundaries. 
The relative positioning of the SC and SG phases
presents a theoretical landmark for comparison with experiments in heavy
fermion systems and high $T_c$ superconductors. 
We find a crossover line traversing the SG--phase 
with $(\mu=0,T=0)$ as its quantum critical (end)point in complete
RPSB, and scaling is proposed for its vicinity.
We argue that this line indicates a random field instability
and suggest Dotsenko--M\'ezard vector replica symmetry breaking
to occur at low temperatures beyond.
\end{abstract}


%% file: intro.tex

\section{Introduction}

The goal of this paper is to present a phase diagram describing the
competition between spin glass order and superconductivity for cases,
which are perhaps best
introduced by the example of Heavy Fermion Systems
(HFS)\cite{mydosh,steglich,barth}. 
Experiments provided evidence for the fact that the same type of
fermions appeared to be responsible for both superconductivity and spin
glass order \cite{steglich}. If one extends this problem of competition
and coexistence between magnetism and superconductivity to include as well
antiferromagnetism for cases of almost absent disorder or of too small
frustration, even more examples for the necessity of single fermion
species models can be found, including high-$T_c$ superconductors
\cite{chou,aharony,scalapino}.\\
We wish to address the competition and coexistence problem of SG versus
SC-ordering in the context of a many fermion model which appears to be
particularly adapted to this kind of problems.\\
Interacting many fermion systems are famous for their complicated
interplay of low--lying excitations of various kinds. Soft breaking of
continuous symmetries and Ward identities are at their origin. 
For the Sherrington-Kirkpatrick Ising spin glass with spin variables
living on Fock space, a similar scenario - even if restricted to
low--lying single fermion excitations -   
may come as a surprise, since the Ising model has only $Z_2$ symmetry and
hence misses the continuous symmetry in spin space,
which guarantees soft spin modes in case of Heisenberg models. 
It is thus important to realize that it needs the full Parisi replica
permutation symmetry breaking (RPSB) as a minimum requirement to
recover the true quantum--dynamics of fermion correlations in the infinite
range fermionic Ising spin glass. Without any
further model--ingredient like fermion hopping or Glauber dynamics for
example, the fermionic Ising spin glass joins static spin behaviour
on one side with a complete quantum--dynamic scenario
in its fermion Green's functions on the other, just as in standard
interacting many--fermion systems. Once the fermionic spin glass becomes a
part of more general model Hamiltonians, these features infiltrate all
coupled degrees of freedom. In this paper, we derive consequences of
this very fact for the case of a competing superconducting instability.\\
On the basis of the proof of soft modes \cite{eplrobr}, calculations of
one-step Parisi RPSB, which already cover more than half of the total
correction obtained from infinitely many steps, are often sufficient to
guess the correct result, i.e. to imagine important features of the exact
solution.
Recalling the crucial importance of the presence or absence of soft modes,
it is clear that once the Ising spin glass with gaussian distributed
exchange interaction is involved in more complicated many fermion models,
it plays a role very different from the pure Ising model. We remark that  
a standard Ising spin glass coupled to a fermionic system by a Kondo
interaction constitutes a different case.\\
The competition between spin glass order and superconductivity has often
been considered to arise as a coupling effect between two systems, one of
which undergoes magnetic order while the other one eventually becomes 
superconducting. 
This issue was addressed by a number of groups in the recent years
\cite{grest,georges,theumann}. 
If one considers metallic conduction instead of superconductivity
one can also find results, obtained along the same lines and with similar
modeling, for example described in the book by Fischer and Hertz
\cite{fischerhertz}. \\
In addition to provide a theory close to the conditions of HFS, we intend
to present our theoretical statements such that a comparison with
phase diagrams of HighTcSuperconductors (HTS) showing a
spin glass phase in between the antiferromagnetic and superconducting ones
\cite{mydosh,chou,scalapino,wbrenig} becomes possible. Of course this
requires to discuss a classification of more or less robust properties
against (low--)dimensional fluctuations.\\
Strontium doped HTS are the most prominent examples,
where antiferromagnetism gives way to clear signatures of spin
glass (SG) like behaviour before superconductivity sets in at higher
(hole)-doping. A typical SG order parameter was
identified at moderately low temperatures \cite{chou} $T<8K$
and even an
infiltration within the superconducting domain at lowest temperatures was
described \cite{scalapino}. Classes of HTS exist as well,
which do neither seem to show spin glass nor intermediate phases.
It is within the scope of our theory to derive conditions and features,
which are specific and in some cases universal. This should provide a
means to identify similar behaviour in real systems.\\
Existence and nature of intermediate spin glass or SG-alike phases in a
certain doping range must be expected for example to be
important features of strongly correlated systems. Experiments
revealing close relations between magnetism and superconductivity in heavy
fermion systems addressed coexistence, phase separation, and
pairbreaking of local pairs by frozen moments for example
\cite{mydosh,steglich}.
\\
Most of recent theories for HTS materials focussed on the
destruction of antiferromagnetism under doping. Viewing an
intermediate phase from the superconducting side, the appealing
concept of a nodal liquid \cite{MPAFisher} was developed.
The role of quenched disorder and randomness was not yet considered, but
its presence should very well participate in the fluctuation destruction
of superconductivity. Since we wish to deal in this paper with
superconducting transitions under participation of a spin glass, a
disorder model is a natural choice.
Arguments on an important role of disorder in HTS and in HF-systems
were provided experimentally and by theoretical reasoning
\cite{mydosh,aharony,steglich}.
Aspects such as non Fermi liquid behaviour, seen to arise in the
vicinity of spin glass order \cite{georges}, also support this point of
view.
\\
In this article we present detailed results for spin glass to
superconductor transitions in a single-species fermionic model, which
treats frustrated magnetic and attractive interaction on the same footing.
Unique features of the phase diagram are derived analytically and
numerically.
The domain of applicability of our model to two-species models, as
recently proposed \cite{theumann} by integrating out conduction electrons
coupled to the fermionic spin glass, depends on the Kondo--effect and
whether the magnetic moments forming the glassy order become quenched or
not. This needs further analysis. We discuss below a related case emerging
in the Periodic Anderson Model in section \ref{pam}. In terms of our
presentation as a single--species model the problem of quenched 
moments appears in form of a metallic spin glass - paramagnet transition. 
\\ 
For the present one--species model we shall observe that for certain
interaction ratios the location of the spin glass bears resemblance to
that of a logarithmic resistivity regime residing
above a spin glass ordered phase at lower $T$ in Sr-doped HTS
\cite{chou,aharony}.
A fluctuational state of broken down spin glass order - perhaps due to low
dimensionality -
should contribute to transport properties seen in intermediate phases
above $T_f$. In particular, SG order was recently shown to affect
transport properties strongly\cite{robrprb}, an effect that
can well have a weak localization precursor due to the random
magnetic interaction.\\
A HTS mechanism related to magnetic fluctuations of a broken down spin
glass appears possible and studies thereof quite justified.
\section{Outline of the paper}
The paper covers the different issues of one--step replica symmetry
breaking in the Sherrington--Kirkpatrick model on Fock space, the
extension of the local theory of disordered superconductivity to arbitrary
filling, and finally the competition between glassy magnetic
order and superconductivity.\\
It is organized in three interrelated larger pieces.\\
I. Several 1--step RPSB solutions for the fermionic Ising spin glass
are presented in section \ref{sg1rsb}. The discovery of a new
random field instability traversing the entire spin glass phase is exposed
by the crossover line Figure \ref{instabline} of section \ref{sg1rsb}.\\
The new 1--step RPSB solutions serve to evaluate the free energies in the
$(\mu,T)$--plane required to obtain the superconductor spin
glass phase diagram in the final part of the paper. \\
II. In two intermediate sections, \ref{sc1} and \ref{scM6}, we report
progress obtained by means of the computer algebra program Mathematica for
the local theory of superconductivity. \\
This local theory adapts to some of the basic conditions of disordered
heavy fermion systems.\\
This part also emphasizes the relation with the $d=\infty$-technique
for clean systems. In these sections, we take explicitly into account
fermion hopping effects, which become dominant in the low temperature
part of the superconducting phase. In the rest of the paper we focus on
superconductivity arising in the magnetic band generated by the
frustrated magnetic interaction, which is also
responsible for spin glass order emerging under favourable conditions.\\
III. Sections \ref{scsg0rsb} and \ref{scsg1rsb} contain the SG - SC
phase diagram. 
We chose to consider the competition problem between SG and SC order
under the condition of a hopping band small in comparison   
with the magnetic band generated by the frustrated magnetic
interaction, which renders corrections from fermion hopping negligible.
Preliminary discussions are given in section
\ref{scsgbasics} and the final result is that of Figure
\ref{fig:diagram1RSB}. We compare this phase diagram with the wellknown
ferromagnet - spin glass diagram (see for example
Ref.\onlinecite{binderyoung}). 
In order to see the relationship, one only needs
to replace first the ferromagnetic- by the attractive interaction and,
secondly, the magnetic field by the chemical potential (note: by means of
a partial particle--hole transformation, one may convert the chemical
potential back into a magnetic field, but then the frustrated magnetic
interaction would turn into a charge interaction reflecting thus the basic
difference). The shape of the SG - SC phase diagram shown in
sections \ref{scsg0rsb} and \ref{scsg1rsb} confirms the genuine 
features.\\ 


%% file: results.tex
\section{Main results} 1. One of the main results and the motivation
of this paper is the derivation of the phase diagram for a fermion
system with competing frustrated magnetic interaction and
superconducting order in sections \ref{scsg0rsb} and \ref{scsg1rsb}.
Our final answers can be found in the Figures 6, 7 (replica--symmetric
approximation) and finally in Figures 11-13 including symmetry
breaking effects.  The phase diagram can be viewed in connection with
the famous classical counterpart of spin glass - (anti)ferromagnet
boundaries \cite{binderyoung}.  Several different features arise, but
a common one to both cases is the suppression of reentrant behaviour
due to replica permutation symmetry
breaking, evidenced by Figure \ref{fig:diagram1RSB}. \\
2. As intermediate steps we derived novel one--step RPSB results for
the order parameters for all temperatures and for a wide range of
chemical potentials by extremalizing the free energy in a
fourdimensional parameter space, as reported here. In a
five--dimensional space we determined
a crossover line indicating\\
3. a new type of random field instability.\\
This part is also new for the pure spin glass problem irrespective of
the competition with superconductivity.  The features of this
instability deserve separate attention in another publication. We
report only those parts needed to resolve the superconductivity
coexistence problem and some features related to the
appearence of a new Quantum Critical Point (QCP).\\
4. Also as intermediate steps to achieve the main goal, we derive a
couple of new results for the theory of superconductivity with local
Wegner invariance\cite{wegner} in sections \ref{sc1} and \ref{scM6}.
This type of superconductivity is marked by a two--particle phase
coherence length playing a very similar role as the usual
one--particle coherence length (the latter one being suppressed by
local invariance which results under
the disorder ensemble average).\\
5. We obtain several exact results and relations for both the
superconductivity and the spin glass issues. We consider as a nice
example the results for the normal and anomalous Green's functions,
expressed in terms of the Kummer function (also known as
Hypergeometric U-function), which display a unique type of spin glass
pairbreaking effect (due to the proximity in the phase diagram). One
may view this also as the effect of
the random magnetic interaction within the superconducting phase.\\
6. With these results we evaluate the complete crossover from BCS to
Bose condensation type superconductivity (see Figure
\ref{bcsbosecrossover}), the crossover being controlled by the ratio of
hopping bandwidth and attractive interaction. Similarities between the
effects which the local Wegner invariance has on disordered
superconductivity and those of the $d=\infty$--technique, usually
applied to the Hubbard model, are remarked.


%% file: model.tex
\section{The model}
\label{model}
We consider here a model described by the grand canonical Hamiltonian
\begin{equation}
{\cal{K}}\equiv{\cal{H}}_{J v}+{\cal{H}}_t-\mu\sum n_i, 
\end{equation}
composed of
\begin{eqnarray}
& &{\cal{H}}_{J v}=-\frac{1}{2}\sum J_{ij}\sigma_i\sigma_j-\sum
v_{ij}a_{i\downarrow}^{\dagger}a_{i\uparrow}^{\dagger}
a_{j\uparrow}a_{j\downarrow},\\
& &
\hspace{2cm}{\cal{H}}_t=\sum_{ij\sigma}t_{ij}a_{i\sigma}^{\dagger}a_{j\sigma},
\label{Hamiltonian_t}
\end{eqnarray}
where $\sigma_i=n_{i\uparrow}-n_{i\downarrow}$,
$a_{i\sigma}$, $n=n_{\uparrow}+n_{\downarrow}$ denote
spin, fermion, and fermion-number operators respectively.
The variance $J^2$ of the frustrated, infinite-ranged and
Gaussian--distributed magnetic interaction $J_{ij}$ and its magnitude
relative to that of the attractive coupling, $v_{q=0}/J$ are relevant
parameters below, together with the chemical potential and the related
filling factor $\nu(\mu)$.  
We do not restrict the attractive interaction to be local, which would
mean a negative $U$ Hubbard interaction. Quantum spin-dynamics would then
exclusively be linked to the fermion hopping $t_{ij}$. 
The mean field approximation may of course be viewed as exact in the
limiting cases of either infinite-ranged $v_{ij}$, infinite number of
orbitals per site, or in the case of infinite dimensions. 
In the latter limit the one particle Green's functions become site-local
in the similar way as in the ensemble average for the present model.
The nonlocal and translationally invariant attractive interaction
$v_{i-j}$, which allows for pair-hopping, is compatible with the
site-local property. We note that local pairing on the average does not
prevent BCS-like behaviour as can be seen for example explicitly from the
discussion below.
The model fits particularly cases encountered in HFS, where one fermion
species appears to be responsible for superconductivity and magnetism.\\
We restrict the discussion to the small $t/J$ regime, which implies
that the selfconsistently determined magnetic band is much larger than
the hopping bandwidth in general. Only deep within the superconducting
regime, where the magnetic bandwidth almost shrank to zero, the single
fermion hopping bandwidth becomes dominant.\\
Local pairs can be delocalized due to finite range $v_{ij}$ or by
arbitrarily weak fermion hopping $t_{ij}$. We employ a local pairing
theory of superconductivity based on the order parameter
$\Delta\equiv\langle a_{i\uparrow}a_{i\downarrow}\rangle$, and on a
two-particle coherence length replacing the usual one-particle length
in the corresponding Ginzburg-Landau theory.


%% file: scsgbasics.tex
\section{The basic selfconsistency equations of the fermionic Ising spin
glass with superconducting order}
\label{scsgbasics}
We begin this paper with the insulating fermionic Ising spin
glass with an additional local pairing order parameter $\Delta$.
The theoretical foundation of the fermionic Sherrington Kirkpatrick Ising
spin glass in the plane of complex chemical potential is quite rich.
Many features had been elucidated in previous papers and thus
only the
important facts needed to understand the present work should be repeated.
In contrast to the standard classical Sherrington Kirkpatrick model, which 
is realized for example at $\mu=i\pi T/2$ in the set of grand--canonical
fermionic Ising spin glasses defined with complex chemical potential, 
the continuous subset on the real $\mu$ axis has two faces: a static one
for all spin and charge--correlations and a quantum--dynamic one for the
fermion correlators like Green's function etcetera. This led us recently
to discover the {\it Quantum--Dynamical Parisi Phase} in this parent model
to many other ones in theory of disordered interacting fermion systems.\\
A rather wellknown and frequently considered source of quantum 
spin--dynamics has been the transverse field in models defined on spin space
\cite{rieger,rieger1,rsy}.  
It should be evident for the reader that the fermionic Ising spin glass, 
complemented with a decoupled attractive fermion interaction to allow for
superconducting order, also develops quantum spin dynamics. 
This holds true after mean--field decoupling of the attractive interaction
with a finite superconducting order parameter $\Delta$, since the
anomalous term does not commute with the Ising Hamiltonian.
There will also emerge quantum dynamical behaviour in charge correlations
and most important, the underlying quantum--dynamical Parisi features of
the single-- and many--fermion propagators are involved too. Thus the
fermionic Ising spin glass with just superconducting order contains
already more physics
than any quantum spin glass in the traditional sense, since the latter
models are all represented on the imaginary $\mu$--axis and their
fermions are not real physical objects.\\
Quantum--dynamical effects are usually dealt with in an approximate way,
a very good analytical idea - depending on the application one has in mind
- being the one of Fedorov and Shender \cite{fedorovshender}, another one
developed by Subir Sachdev for Ginzburg Landau theories \cite{rsy}
and employed for the metallic spin glass too \cite{rso}. Unlike the
Heisenberg model, where Usadel proved the importance of spin dynamics for
the absence of replica-symmetric domains within the spin glass phase
\cite{usadel}, spin--dynamical corrections to the Ising spin glass -
superconductor boundary are negligible. On the contrary 
replica symmetry breaking is linked to the quantum fermion
dynamics and turns out to be essential at low temperatures.
Our results presented in section \ref{scsg1rsb} show that it
suppresses reentrant SG - SC - SG transitions.
The latter point of course recalls what has
been found for the ferromagnetic - spin glass boundary in the
Sherrington--Kirkpatrick model \cite{binderyoung}. \\
The effects of Parisi's RPSB complicate the appearance and the treatment
of the selfconsistency equations such that it is necessary to start with
the unbroken simpler ones. Already these equations turn out to have
multiple solutions, whose stability changes eventually as a function of
chemical potential and temperature. The stable solutions must first be
understood before the one step RPSB analysis can be started.
\subsection{The free energy and resulting selfconsistency
equations in replica symmetric approximation}
The grand canonical model ${\cal{H}}_{Jv}$ can be decoupled in the
standard way.
The Grassmann field theory depends on how many steps of replica symmetry
breaking are taken care of - hardly necessary to say that one can rarely
hope to find full analytical solutions. Already one--step breaking is
nonsimple and in case of the fermionic model is almost as complicated as
two--step breaking in the standard case on spin space. 
Let us start with the free energy obtained in the replica symmetric and
$Q$--static approximation as
\begin{equation}
f=\frac{1}{4}\beta J^2
((\tilde{q}-1)^2-(q-1)^2)+\frac{|\Delta|^2}{v}
- T \ln 2 -\mu -T\int_z^G \ln{\cal{C}},
\label{fe0rsb}
\end{equation}
where
\begin{equation}
{\cal{C}}(z)\equiv
\cosh(\beta\tilde{H}(z))+\cosh(\beta\sqrt{\mu^2+|\Delta|^2})
\exp(-\frac{1}{2} (\beta J)^2
(\tilde{q}-q)) \quad \mbox{and} \quad \tilde{H}(z)\equiv J \sqrt{q} z
\label{calC}
\end{equation}
Here, we have introduced the convenient short--hand notation for the
Gaussian integral operator
\begin{equation}
\int_z^G = \frac{1}{\sqrt{2 \pi}} \int_{- \infty}^{\infty} dz
\exp(-z^2/2).\nonumber
\end{equation} 
The dynamic susceptibility $\chi(\omega)$ is approximated by the
$\omega=0$--part denoted by $\tilde{q}\equiv\chi(\omega=0)\hspace{.1cm}T$,
an approximation which becomes exact however for vanishing superconducting
order parameter. Thus the equations are sufficient to determine the
replica symmetric phase boundary. Even in case of discontinuous
transitions, dynamic corrections are very small, first as long as the jump
of superconducting order parameter rests small, second because the free
energy integrates over the dynamic effects of Lorentzian broadening
\cite{fedorovshender}. No qualitatitive change can thus be expected from
this type of dynamics. Deeper inside the superconducting phase the
calculation with the static $<\sigma_i(t)\sigma_i(t')>$ becomes
approximative but still interesting. The exact mean field  
solutions are much more involved than those known from the
pairbreaking effects of random magnetic impurities, the latter case being
included in section \ref{scM6}. We hope to come back
to the dynamic corrections around the new solutions so far obtained in
$\chi(\omega)\approx \chi(0)=\beta\tilde{q}$ approximation.\\ 
As the superconducting order increases the susceptibility $\chi$ is
depressed strongly and using the static approximation for $\chi(\omega)$
again does not deteriorate the treatment. 
All this will be worked out below in the section \ref{sc1} of local
pairing superconductivity with the local Wegner - invariance.\\
Comparing with the transverse field Ising model, it is here the
superconducting order parameter - or, if
one looks at the original model, the attractive interaction - which
introduces quantum dynamics on the level of spin and charge. Ignoring
quantum phase transitions the $\omega=0$ approximation is sufficiently
good and an improvement can be obtained perturbatively with a Lorentzian
form originally derived by Fedorov and Shender for the transverse Ising
spin glass \cite{fedorovshender}.\\ 
It is standard knowledge that, due to the replica limit and in agreement
with the search for local stability, one must maximize the free energy
functional with respect to the n(n-1)/2 off--diagonal matrix elements
$q^{a\neq b}=Q^{a\neq b}|_{s.p.}$ but minimize with respect to the n
diagonal ones $q^{aa}=Q^{aa}|_{s.p.}$. With growing number of variational
parameters - five at one step RPSB in the fermionic case (only three in
the standard spin model) - this becomes a tedious numerical problem.\\
Extremizing the free energy in the given lowest order form for the
coupled spin glass - superconductor problem leads to the
following set of coupled replica--symmetric selfconsistent equations,
where we from now on set $J=1$, i.e. we measure all energies ($f$,
$T$, $\mu$, $\Delta$, and $v$) in terms of $J$.
\begin{eqnarray}
q&=&\int_z^G \sinh^2(\beta\tilde{H}(z))/{\cal{C}}^2(z)\\
\tilde{q}&=&\int_z^G \cosh(\beta\tilde{H}(z))/{\cal{C}}(z)\label{scsgqt}\\
m&=&\int_z^G \sinh(\beta\tilde{H}(z))/{\cal{C}}(z)\\
\nu&=&1+\sinh(\beta\sqrt{\mu^2+|\Delta|^2})\exp(-\frac{1}{2} \beta^2
(\tilde{q}-q))\int_z^G {\cal{C}}^{-1}(z)
\label{sceqs0rsb3}\\ 
\Delta&=&\frac{v}{2}\Delta
\frac{\sinh(\beta(\sqrt{\mu^2+\Delta^2}))}{\sqrt{\mu^2+\Delta^2}}
\exp(-\frac{1}{2}\beta^2(\tilde{q}-q))\int_z^G{\cal{C}}^{-1}(z)
\label{sceqs0rsb4}
\end{eqnarray}
A mapping of the pure Ising spin glass with chemical potential $\mu$ to
the half--filled model with order parameter $\Delta$ is obvious. That's
why the tricritical point derived at $(\mu=.961 , T =1/3)$ for the Ising
SG \cite{brro} without superconductivity must be present in the
half--filled spin glass superconductor.\\
The selfconsistent equations (\ref{scsgqt}) and (\ref{sceqs0rsb4})
can be combined into  
\begin{equation}
\tilde{q} = 1 -
2\frac{\sqrt{\Delta^2+\mu^2}}{v}\coth(\beta\sqrt{\Delta^2+\mu^2}),
\quad\Delta\neq 0,
\label{qtofdeltarelation}
\end{equation}
which holds in the superconducting phase. The spin glass order parameter
does not appear explicitly in the relation between superconducting order
and spin autocorrelation function. Relation (\ref{qtofdeltarelation})
can be used to express both susceptibilities, equilibrium ($\chi$)
and nonequilibrium ($\bar{\chi}$) as well. \\ 
One may combine the relations (\ref{sceqs0rsb3}) and
(\ref{sceqs0rsb4}) for $q=0$ to realize that $\nu$ does not depend on $
\mu$ below the SC transition, which is a typical feature of Bose
condensation type superconductivity.
This reflects the fact that the above equations neglect fermion
hopping within the $O((t/J)^0)$ approximation. As section \ref{sc1} shows,
see also Figure \ref{bcsbosecrossover}, the introduction of strong
enough fermion hopping Eq.(\ref{Hamiltonian_t}) leads to a smooth
crossover to BCS--type superconductivity. 
The ratio of hopping bandwidth versus
magnetic bandwidth generated here by the frustrated Ising interaction
decides which type of superconductivity is obtained. We focus in this
paper on superconductivity in the magnetic band, neglecting whenever
possible the influence of an additional hopping band for the same type
of fermions.\\
In the final discussion of section \ref{pam} we stress the transition from
a two--band spin glass for zero or very small hopping to
an effectively a three--band case, where a central hill of the
fermionic density of states represents the scattering into nonmagnetic
states.  We emphasize that pairing in this nonmagnetic part may lead
to a different answer on the coexistence of spin glass order and
superconductivity. This is not analysed in the present paper and
requires a separate analysis.\\  
We are now in a position to proceed with the question 
of superconductivity in the presence of spin glass order.\\ 
The reader who is neither interested in replica symmetry breaking nor in
details of the superconductivity theory, which involve local one--particle
Green's functions in the ensemble--average, may jump to section
\ref{scsg0rsb}, where the SG - SC coexistence is analysed which ends
up in the phase diagram for the whole $(\mu,T)$--plane. The important
corrections of replica permutation symmetry breaking,
worked into the final phase diagram of section \ref{scsg1rsb} should
not be missed.


%% file: sg1rsb.tex
\section{Replica permutation symmetry breaking solutions in
the plane of all temperatures and chemical potentials}
\label{sg1rsb}
\begin{figure}
\centerline{\epsfbox{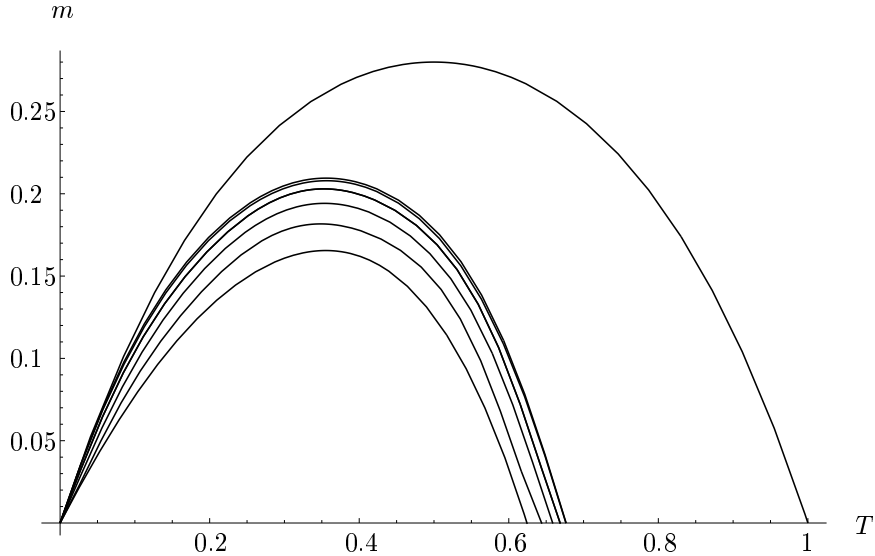}}
\caption{A selection from our results for the Parisi parameter $m(T)$ at
various chemical potentials $\mu=0, .1, .2, .3, .4$, and $.5$, identified
by their right endpoints $(T_c(\mu=0)=.6767,m(T_c)=0), (.6765,0),
(.669,0), (.659,0), (.6444,0)$, and $(.6247,0)$
respectively , shown in
comparison with the spin space result of Parisi \cite{parisi80b}
(endpoint:$(1,0)$),
which corresponds to $\mu=i\frac{\pi}{2}T\hspace{.1cm}\mathrm{mod}(2\pi
T)$ in the Fock space
model and is different from half--filling.} 
\label{m}
\end{figure}

\begin{figure}
\centerline{
\epsfbox{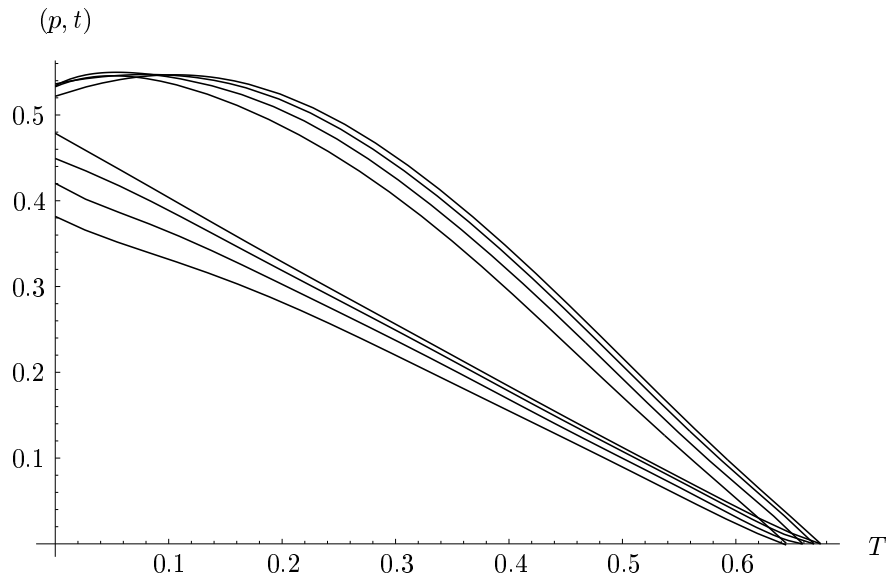}}
\caption{Parisi order parameters $p$ and $t$, $q_1=p+t$, $q_2=p$ for
$\mu=.1, 
.2, .3$, and $.4$; the upper bundle of curves, showing maxima away from
$T=0$, refer to $t$, the lower bundle to $p$; $\mu$--values on individual
curves are identified by the critical endpoints as in Fig.\ref{m}} 
\label{pandt}
\end{figure}

\begin{figure}
\centerline{
\epsfbox{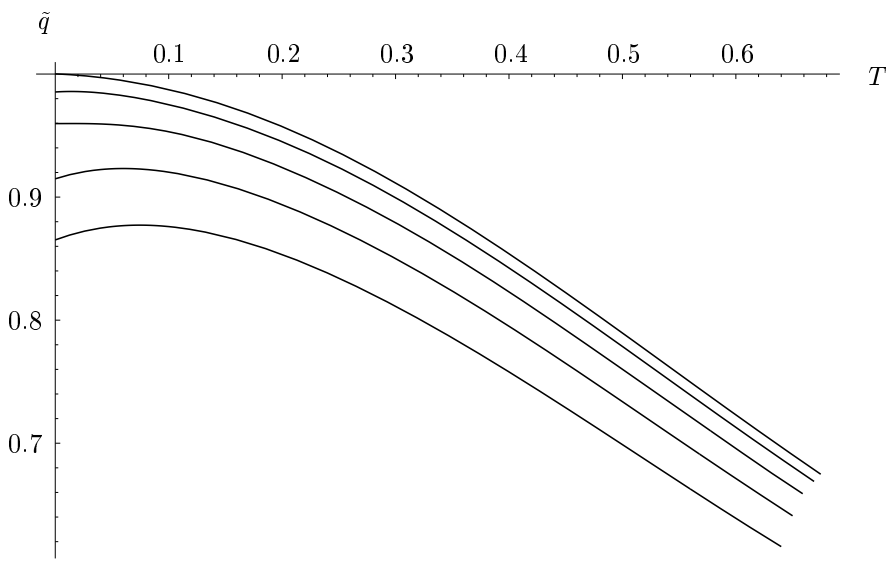}}
\caption{The spin autocorrelation in the fermionic Ising spin
glass phase, shown for $\mu=.1, .2, .3, .4$ (from upper to lower
curve, respectively); ($\tilde{q}(\mu=i\pi T/2)=1$
is the standard SK--value on spin space). $\tilde{q}(T\rightarrow 0)<1$
beyond half--filling.} 
\label{qt}
\end{figure}

Locating the phase boundary between spin glass and superconductivity
within the entire $(\mu,T)$--plane requires to solve a couple of
entangled problems:
as a consequence of the multiplicity of solutions (even in the
paramagnetic regime) and of discontinuous nature of the transition, 
both a thorough investigation of the set of solutions to the
selfconsistent equations and finally the evaluation and comparison of 
their free energies was required. This complication arose already without
replica symmetry breaking. 
\\
In addition, as for the
ferromagnet-SG transition \cite{binderyoung}, replica symmetry breaking is
important and corrections need to be controlled:
It turns out that superconductivity repells a
considerable part of the magnetic domain down to lowest temperatures. The
lower the temperature is the stronger are replica symmetry breaking
effects. For this reason we first generalize the replica symmetric
analysis presented in the last section to one-step breaking. 
This ensemble of solutions for the 1RPSB Parisi solution of the fermionic
Ising spin glass, needs to be presented here as a basis
for our phase diagram analysis. \\ 
We shall see that in many respects 1step RPSB gives enough information to
guess the exact solution. 
The analysis at one step breaking requires however some
preparation from the pure fermionic spin glass problem which we shall
provide now.\\
The free energy per site at 1RPSB is given by
\begin{equation}
f = \frac{\beta}{4}\left[(1-\tilde{q})^2-(1-q_1)^2+
m(q_1^2-q_2^2)\right]- T \ln2 - \mu + \frac{|\Delta|^2}{v} - \frac{T}{m} \int_{z_2}^G 
ln \int_{z_1}^G {\cal{C}}^m 
\end{equation}
where ${\cal{C}}$ is the same expression as Eq.(\ref{calC}) however
with $q$ replaced by $q_1$ and the 1-step RPSB broken effective field 
\begin{equation}
\tilde{H}(z_1,z_2)=\sqrt{q_2}\;z_2+\sqrt{q_1-q_2}\;z_1
\end{equation}
One can either use $p$ and $t$ in the Parisi step--height--notation
\cite{parisi80b} or the two order parameters
$q_1=p+t$ and $q_2=p$ as variational parameters together 
whith the third order parameter $\tilde{q}$ as the static
saddle point of the $Q^{aa}$--field - alternatively
one may choose the linear equilibrium susceptibility,
the Parisi parameter $m$, and the superconducting order parameter
$\Delta$.
This is a variational problem in five--dimensional space (comparable to
2--step RPSB in the standard model).
The free energy must be simultaneously maximized with respect to
$p$, $t$, $m$, and minimized in $\tilde{q}$ and $\Delta$. Since it turns
out that a coexistence phase in the sense of $q\neq 0,\Delta\neq 0$ does
not exist except for nonzero magnetic fields below the upper critical
field strength, there is no meaning in showing the Parisi solutions for
finite $\Delta$. Hence let us infer (details will be published elsewhere)
our numerical solutions for $\Delta=0$. All quantities are derived down to
lowest temperatures, supplemented and aided by exact analytical
relations which are given. The Parisi parameter $m$ is shown in Figure
\ref{m}, the order
parameters $p$ and $t$ are displayed in Figure \ref{pandt}, and the spin
autocorrelation function $\tilde{q}$ is shown by Figure \ref{qt}.

The curves differ more and more as the chemical potential increases; we
are not showing curves at $\mu$--values close to the thermodynamic first
order transition, since they suffer from a higher order instability
which we evaluate below.

The linear equilibrium susceptibility is given by the relation
\begin{equation}
\chi=\beta(\tilde{q}-(1-m)(p+t)-m p)
\end{equation}
and the fermion filling factor obeys exactly (to all orders of RPSB)
\begin{equation}
\nu=1+\tanh(\beta \mu)(1-\tilde{q}) 
\label{nuandqt}
\end{equation}
The filling factor, being a first derivative of the free energy with
respect to the chemical potential, has a slope which is connected to 
$\partial^2 f/\partial\tilde{q}^2$.
\begin{equation}
\frac{\partial^2
f}{\partial\tilde{q}^2}=\frac{\beta}{2}[1+
\frac{\beta}{2}\coth^2(\beta\mu)\{\frac{\beta(1-\tilde{q})}{\cosh^2(\beta\mu)}-
\frac{\partial\nu}{\partial\mu}\}]
\label{relation}
\end{equation}
The latter equation turned out to be related to the absence of simple
solutions for large enough chemical potentials and low enough
temperatures. Since a direct extension of the AT analysis is hampered by complex replica--diagonal AT eigenvalues, the zeros of eq. \ref{relation} were suggested in a related problem as signs of a new instability \cite{lage82a}. Other authors\cite{mottishaw85a} demanded a negative real part of the replica--diagonal eigenvalue to be necessary for this instability. The line of zeros of eq. \ref{relation}, however, has the virtue that some of its properties can be calculated for complete RSB.
All dependence on the spin glass order parameters is absorbed in
$\partial\nu/\partial\mu$, which is proportional to the
thermodynamic density of states at the Fermi level. 
Since Eq.(\ref{relation}) holds for arbitrary
numbers of RPSB--steps and since the spin autocorrelation function cannot
reach the saturation value $\tilde{q}=1$ away from half--filling, one
would need a pseudogap at the Fermi level to allow for a positive value of (\ref{relation}).
at zero temperature. This would however be necessary for any value of
$\mu$, which in turn would imply that one cannot reach a filling factor
different from 1 (half--filling). Thus a pseudogap assumption for the
thermodynamic density of states would not cure the problem.
Hence it seems to us that vector replica symmetry breaking of the
$\tilde{q}$ should be considered. The landscape of the free energy
functional around the point where the minimization in $\tilde{q}$ is lost,
while the common maximization w.r.t. the other quantities is maintained,
at first sight only suggests that an integration over the range of the
field $Q^{aa}$ might be necessary. In the present paper we restrict this
analysis to the determination of the line 
$\frac{\partial^2}{\partial\tilde{q}^2}f=0$, which locates the crossover
to a regime, where f is not minimized by $\tilde{q}$ but
(replica--diagonal) AT--stability did not yet break down.
More precisely, the divergence of the replica--diagonal susceptibility
$\langle Q^{aa}Q^{aa}\rangle$ does not yet occur on the crossover line
because of the finite coupling to the noncritical $Q^{a\neq b}$--fields.
Since the latter are massive inside
the SG phase, integrating them out leads to a shift of the critical
point, and the divergence occurs on the (replica--diagonal) AT instability
line below.
This leaves unchanged the fact that it is a pure random--field
critical phenomenon. Of course $Q^{aa}Q^{bb}$--couplings are generated by
the elimination of the noncritical fields $Q^{ab}$.

\subsection{Random field instability in the low
temperature region of the spin glass phase}
\begin{figure}
\centerline{
\epsfbox{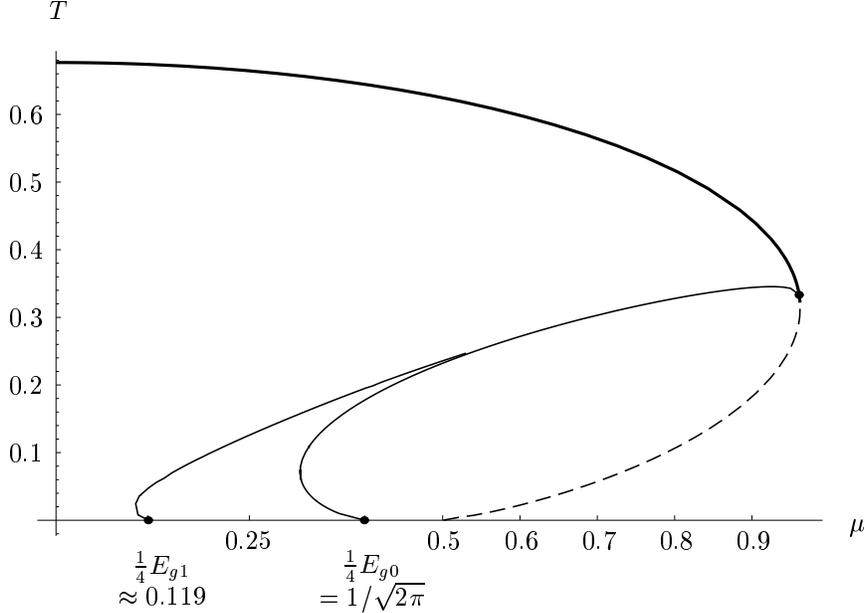}}
\caption{The spin glass phase diagram as a function of chemical potential
$\mu$ and temperature $T$. The line $\partial^2
F/\partial\tilde{q}^2=0$ is
shown in replica symmetric approximation and in one step symmetry
breaking. The 0RPSB--line hits the $T=0$--axis exactly at
$\mu_0=\frac{1}{\sqrt{2\pi}}$,
while the 1RPSB--line progresses to $\mu_1\approx .1195$ at $T=0$,
both values being equal to one half of the 0RPSB - and 1RPSB -
gap energies ($E_{g0}, E_{g1}$) of the fermionic density of states.
The 1RPSB--line becomes almost identical with the 0RPSB--line for
$\mu>.6$. The exact solution at $\infty$--RPSB ends at
$(T=0,\mu=E_{g\infty}=0)$.} 
\label{instabline}
\end{figure}

Our numerical analysis of the crossover line, determined by
$\frac{\partial^2}{\partial\tilde{q}^2}f=0$,
was carried out through the entire phase diagram. Its high $T$ end is
always the SG--tricritical point \cite{brro}, while its zero
temperature endpoint is given by $\frac{1}{4}$ of the fermion density
of states gapwidth \cite{eplrobr}.  This gap extends from $-E_g$ to
$+E_g$, where $E_g=\sqrt{2/\pi}$ in the replica-symmetric
approximation and shrinks as the number $k$ of steps of Parisi replica
permutation symmetry breaking is increased. For the exact solution the
gapwidth becomes zero with some implications described in
\cite{eplrobr} and the instability line of a vanishing mass of the
$Q^{aa}Q^{aa}$--propagator also follows into the origin. Details of
the calculations and of our finding that the point at half-filling and
zero temperature represents a new quantum critical point are presented
elsewhere. \\
Since in replica symmetric approximation the AT-eigenvalue belonging
to replica-diagonal fluctuations can be expressed in terms of second
derivatives of the free energy \cite{AT}, and since $\partial^2
f/\partial q \partial \tilde{q}$ does not vanish on $\partial^2
f/\partial \tilde{q}^2=0$ except for $T\rightarrow 0$, the
AT--instability line, determined from the zeros of the real 
part of this eigenvalue, lies below the crossover line.\\
The position of the AT--instability line in replica symmetric
approximation can be inferred from the work of da Costa et al
\cite{daCosta94a}, since mapping of the selfconsistent equations to
their $S=1$--model has been demonstrated \cite{brro}. In turn, our
RPSB analysis can be applied to the analysis by da Costa et al of the
$S=1$ Ghatak--Sherrington model.  Instead of the gap energies $E_g$
one should then connect the $T=0$ instability points at each order of
RPSB to the corresponding values of
the nonequilibrium susceptibility.\\
Since the AT-eigenvalue assumes a much more complicated form under one
or more RPSB--steps, we limit the present discussion to the crossover
line.  It has the virtue of allowing for some exact results, formally
independent
of the replica--offdiagonal order parameters.\\
The one--step RPSB correction presented here in Figure
\ref{instabline} illustrates how the random field instability line
(linked closely to the crossover line, both lines joining at the
endpoints) progresses towards half--filling at $T=0$, reaching it
finally under infinite RPSB.

The physical origin of the instability lies in the dilution of
the effective spin density as the particle pressure increases with
$\mu$ (starting from $\mu=0$). Although the random magnetic interaction
wants to magnetize all sites at $T=0$ (and to order them randomly)
this is not possible because of doubly occupied sites. At any finite order
k of k--step replica permutation symmetry breaking there is a finite
charge gap, which
prevents deviations from half-filling until the chemical potential reaches
half of the gap--edge value. Beyond this value, the fermion filling factor
differs from 1 and both the replica symmetric and the one-step broken
solution maximize the free energy instead of minimizing as is required
with respect to $\tilde{q}$. 
The resolution is still an open problem in replica theory at $T=0$. 
We hope to come back in a future publication with the analysis of 
Dotsenko--M\'ezard vector replica symmetry breaking (VRSB) \cite{dotsmez}
which we consider a potential candidate for resolving this problem.
Breaking of translation invariance of the disorder ensemble by means of
instanton solutions can also not be excluded at present.\\
For the purpose of this paper it is sufficient to restrict the discussion
to the region above the instability line (the presented crossover line
presents an upper bound for the instability region), which stretches from
$(T=0,\mu=0)$ to the SG tricritical point in the exact Parisi solution.
We shall see that, for a large interval of interaction ratios $v/J$,
all of the spin glass -
superconductor boundaries derived for the whole range of interaction
ratios fall into the stable regime above the crossover line. 
In addition to the fundamentally important possibility of VRSB
emerging together with Parisi's RPSB, the solution below the instability
line can be fascinating too, since quantum critical phenomena in many
models that are related to the fermionic Sherrington--Kirkpatrick
model are affected.\\
We note in passing that the crossover from stable (F-minimizing) to
the spin glass stabilized regime of $\tilde{q}$--solutions, and the
random--field critical behaviour is hardly
observable in the lower order physical observables like the linear
susceptibility $\chi$.
It is also remarkable that the analogy between the replica--diagonal
vector--like spin glass field $Q^{aa}$ and the (replicated) magnetization
$m^a$ does not hold w.r.t. the instability: the ferromagnetic transition
prevents a negative mass, but the spin glass transition does not. This is
one reason to attempt the Dotsenko--M\'ezard VRSB in the latter case, in
addition to the fact that random--field features considered by Dotsenko
and M\'ezard are present too.\\
We mention in this context that a previous numerical analysis of the
fermionic TAP--equations, invented by us to enable a replica--free
analysis of the fermionic models, showed a central mountain in the density
of states, which belongs to nonmagnetic sites away from half-filling, and
probably as an effect of this the filling factor $\nu(\mu)$ increased
continuously \cite{rehker}.  
\subsection{Fermion filling solution $\nu(\mu,T)$ along the crossover
line and quantum scaling near half--filling due to 
the $\infty$--step RPSB Quantum Parisi Solution}
By means of the two exact relations (\ref{nuandqt}) and (\ref{relation}) 
we are able to describe the crossover line as the solution of the
differential equation
\begin{equation}
\frac{\partial\delta\nu}{\partial\mu}=\frac{2\beta}{\sinh(2\beta\mu)}\delta\nu
+2\hspace{.1cm}T\hspace{.1cm}\tanh^2(\beta\mu),\quad\quad\delta\nu\equiv
\nu-1
\label{instabDGL}
\end{equation}
together with the selfconsistent equation for the filling factor, which
involves all order parameters. 
Even without explicit knowledge of the low temperature
Quantum Parisi Solution (QPS) we can solve for the fermion
concentration along the crossover line, finding
\begin{equation}
\nu(\mu,T) = 1 + \tanh(\beta \mu) (C + 2 T^2 \ln \cosh (\beta \mu))
\end{equation}
where the constant $C\approx 0.18$ is determined at the tricritical point.
Near the limit of zero temperature and of zero chemical potential
(half--filling) the solution carries information of the QPS. Supposing
that the overshooting of the curve near the gap--size related
endpoints at $E_{g,k}/2$ persists, although it
becomes smaller and smaller with increasing order $k$ of RPSB, 
this implies the scaling law
\begin{equation} 
T\sim \mu^\psi
\label{Tmuscaling}
\end{equation}
with a shift--exponent $0<\psi<1$, which produces the infinite
slope of the crossover line near the QCP $(0,0)$. Conclusion
(\ref{Tmuscaling}) also follows from the assumption of a continuous
filling factor as $(0,0)$ is approached on the crossover line. 
Thus, by the use of (\ref{Tmuscaling}) one finds near half--filling that 
\begin{equation}
\delta\nu(\mu)|_{_{\mu\sim T^{1/\psi}}}\approx C\hspace{.1cm}
\mu^{1-\psi}.
\end{equation}
Together with the behaviour\cite{eplrobr} of the single--particle density
of states $\rho(E)$
these are the first results on the scaling behaviour near the Quantum
Critical Point $(\mu=0,T=0)$, which is marked by $\infty$--step RPSB and
perhaps by VRSB in addition.


%% file: sc1.tex
pap\section{Local superconductivity theory for heavy fermion systems
with disorder}
\label{sc1}
Several years ago one of us imposed Wegner's local gauge invariance, which
reflects statistical phase cancelations in disorder ensembles, to define 
a theory of disordered superconductivity. This symmetry resulted in local
one particle Green's functions and hence in a local pairing theory for the
ensemble average. In this same sense superconductivity was based on
two--particle coherence length.\\
While a few central properties are reviewed briefly, omitting as far as
possible repetitions from a series of earlier publications, new results
are given in detail below. Within the scope of the present paper they
serve to study the effects of a random magnetic interaction.   

\subsection{The local superconducting one particle Green's function
solutions at arbitrary filling} 
The selfconsistent equations, realized as exact solutions of
the large n limit in n orbital model formulations but justified also as 
mean--field equations valid for arbitrary n, read

\begin{eqnarray}
& &{\cal{G}}(\epsilon_l)={\cal{G}}_0(\epsilon_l)\left[1-\Delta
{\cal{F}}^{\dagger}(\epsilon_l)+\frac{1}{4}w^2({\cal{G}}(\epsilon_l)^2-
{\cal{F}}(\epsilon_l)
{\cal{F}}^{\dagger}(\epsilon_l))\right]\\
&
&{\cal{F}}(\epsilon_l)={\cal{G}}_0(\epsilon_l)\left[\Delta^*{\cal{G}}
(-\epsilon_l)+\frac{1}{4}w^2{\cal{F}}(\epsilon_l)({\cal{G}}(\epsilon_l)
+{\cal{G}}(-\epsilon_l))\right]
\end{eqnarray}

where $\epsilon=(2n+1)\pi T$ and $w^2\equiv\frac{4}{N}\sum_j\langle
t_{ij}^2\rangle$ determines the hopping bandwidth $2w$.\\ 
The above equations for a time--reversal invariant superconductor were
solved for half filling in \cite{RO}. In the present model the filling
dependence or $\mu$--dependence respectively is very important for the
competition between magnetic order and superconductivity. 
This revived interest in the filling dependence led to
the surprising insight that the equations can also be solved analytically
and controlled for arbitrary chemical potential and filling respectively.
This was unexpected because the equations superficiously appeared to be
of higher than fourth order. 
As usual one can transform away the order parameter phase. 
The Bogoliubov mode needs not be considered in the present context 
of one particle functions (it is anyway respected by the Ward identity
for charge conservation).
The phase transformation effectively reduces the order of the equations
from 6 to 4.
The most convenient form of the one particle Greens functions takes the
form

\begin{equation}
{\cal{G}}(\epsilon_l)=
\frac{i\epsilon_l+\mu}{w^2/2}-\frac{\sqrt{2}}{w^2}\frac{1}{\sqrt{\epsilon_l^2
+\Delta^2}}\sqrt{4i\mu\epsilon_l(\epsilon_l^2+\Delta^2)-
(w^2+\epsilon_l^2+\Delta^2-\mu^2)(2\epsilon_l^2+\Delta^2)+\Delta^2
r_1(\epsilon_l)}
\label{hopping_G}
\end{equation}

and

\begin{equation}
{\cal{F}}(\epsilon_l)=-\frac{\Delta}{2
w^2}+\frac{\Delta}{\sqrt{2}w^2}
\frac{\sqrt{w^2+\epsilon_l^2+\Delta^2-\mu^2+r(\epsilon_l)}}{\sqrt{\epsilon_l^2+
\Delta^2}},
\label{hopping_F}
\end{equation}

where 
\begin{equation}
r_1(\epsilon_l)\equiv\sqrt{-4
w^2\mu^2+(w^2+\epsilon_l^2+\Delta^2+\mu^2)^2}.\nonumber
\end{equation}
All simple limits of half--filling, $\mu\rightarrow 0$, vanishing order
parameter, $\Delta\rightarrow0$, vanishing bandwidth, $w\rightarrow0$,
and $\epsilon_l\rightarrow0$ are correctly reproduced. The analytical
properties in the complex plane are easily analysed. 

In Ref. \onlinecite{RO} it was mentioned that the crossover
from BCS-type to Bose condensation like superconductivity is observed as
the bandwidth/order parameter ratio decreases. As superconductivity,
unlike magnetism, reacts rather weakly to changes of the chemical
potential, this crossover will show up again under arbitrary $\mu$.
The existence of a superconducting current was also studied in detail; the
locality of the disorder-averaged one particle Greens functions does not
prevent ths superconducting behaviour based on two particle hopping
processes.\\
For the moment the disorder induced superconducting glass order parameters
are neglected as small effects, but under strong disorder they can be
important and we extend below the present analysis accordingly. Since our
goal is to analyze quantum phase transitions and in particular the
transition between antiferromagnetic order and/or spin glass order on the
weak filling side and superconductivity on the other, we will naturally
be concerned in the following sections with the question of replica
symmetry breaking at the border and in the bulk of the superconducting
phase. The 
unique effects of ergodicity breaking and aging related to RPSB are
significant and could be experimentally used in high- $T_c$ superconductors 
and heavy fermion superconductors.\\ 
The superconducting order parameter and the filling factor, calculated
according to
\begin{equation}
\Delta=v\hspace{.1cm}T\sum_l {\cal{F}}(\epsilon_l)\quad {\rm and}\quad
\nu=T\sum_l
{\cal{G}}(\epsilon_l)e^{i \epsilon_l 0_+}
\end{equation}
are shown as a function of the chemical potential in
Fig(\ref{bcsbosecrossover}). 

Before we generalize the analysis to coexisting and competing spin glass
order, which forces one to study the possible indirect effect of
RPSB on superconductivity, we study the filling dependence in
presence of two pairbreaking effects considered in Ref. \onlinecite{RO}
for the special case of half-filling.


%% file: scM6.tex
\newcommand{\Mvariable}{}

\section{Exact solution of the local theory with inhomogeneous
superconducting order parameter at half filling}
\label{scM6}
\subsection{Field theory of the decoupled superconductor exposed to
  arbitrary chemical potential} This section serves the purpose to
demonstrate the basic difference between standard pairbreaking from
paramagnetic impurities scattering and the one generated by random
many body interaction and competing spin glass order. Aided by
Mathematica, we
were able to extend earlier solutions \cite{RO}.\\
The complications due to a superconducting order parameter which is
allowed to fluctuate statistically in modulus and in its phase are
considerable away from half-filling. The selfconsistent calculation of
these statistical fluctuation remains difficult and may require random
attractive interactions, while in the case of mesoscopic
superconductors these fluctuations occur as an effect of
nanostructuring \cite{zirnbauer}.
\\
We find that random fluctuations of $|\Delta|$ play the same role as
paramagnetic scattering, while statistical fluctuations of $\phi$ in
$|\Delta|e^{i\phi}$ are unessential.  Let us consider the same set of
second moments as in Ref.  \onlinecite{schohe},
generalize and solve the field theory in saddle-point approximation.\\
The replicated partition function $\langle Z^N
\rangle=\int{\cal{D}}[\Phi]e^{\cal{A}}$ expressed in a grassmann field
theory with the action ${\cal{A}}$ was given in Ref.
\onlinecite{schohe} and will not be repeated here.  The physical
understanding will be sufficiently supported by recalling that the
four different scattering rates
\begin{equation}
\tau^{-1}_t, \tau^{-1}_s, \tau^{-1}_{|\Delta|},\quad {\rm
and}\quad\tau^{-1}_{\phi}
\end{equation}
are referring to nonmagnetic scattering (index t), paramagnetic spin flip
scattering (s), and scattering from statistical fluctuations of phase
($\phi$) and modulus ($|\Delta|$) of the order parameter
$\Delta=|\Delta|\exp(i\phi)$ respectively. No limitations on the size of
these different scattering processes had to be assumed.
All scattering rates are as usual related by
$\tau^{-1}_{\alpha}=2\pi\rho_F M_{\alpha}$ to the corresponding second
moments of the tight binding model. 
\subsection{Exact solution at half-filling}
\begin{figure}
\epsfbox{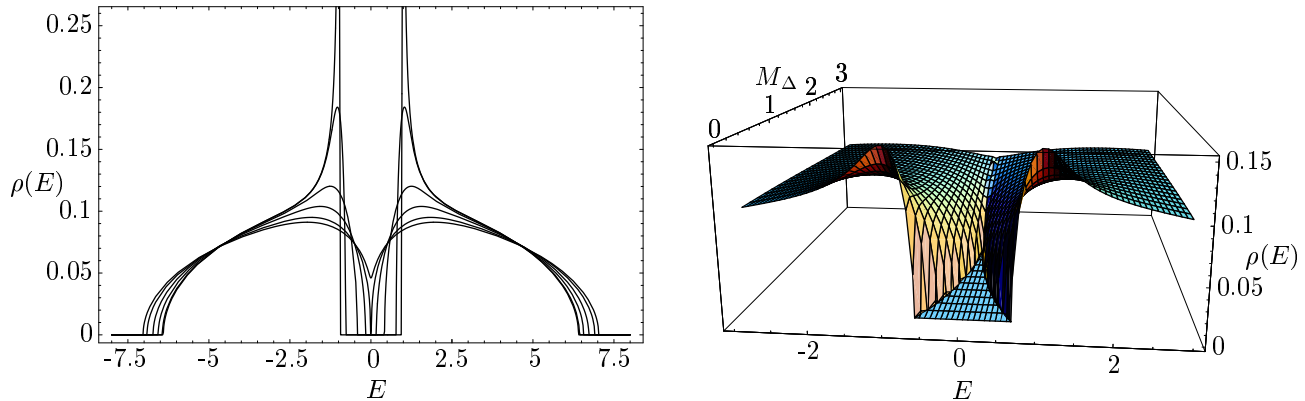}
\caption{Density of states $\rho(E)=- \frac1{\pi} \mathrm{Re} G^R(E)$ with different
values of $M_\Delta$. On the left side, the gap is reduced and finally closed
as $M_\Delta$ runs through $M_\Delta=(0.01,\;0.1,\;0.5,\;1,\;1.6,\;2)$.
The figure on the right side gives an overview for $0.2 < M_\Delta < 3.0$.}
\label{dosgapM6}
\end{figure}

The saddle point equations derived from the field theory \cite{schohe} 
reduce effectively to quartic order at half filling and can
hence be solved exactly. The solution for the Greens function $\mathcal{G}$ 
and the anomalous propagator $\mathcal{F}$ can be written as 
\begin{align}
{\mathcal{G}}(y) =& {\frac{i}{{M_a}\,{M_b}}}\,\left(
{\frac{1}{4}}\,y\,
  \left( 2\,{M_a} + {M_b} \right)  - 
    {\frac{1}{2}}\,{\sqrt{{\frac{{S_2} + {S_3}}{3}}}} - 
    {\frac{1}{2}}\,{\sqrt{-{\frac{{S_2}}{3}} + {\frac{2\,{S_3}}{3}} - 
         {\frac{{\sqrt{3}}\,{S_4}}{4\,{\sqrt{{S_2} + {S_3}}}}}}} 
         \right)\\
{\mathcal{F}}(y) =& {\frac{{\mathcal{G}}\,\Delta }{i\,y +
{\mathcal{G}}\,{M_b}}}
\end{align}
with the definitions $y\equiv \epsilon_l$ and
\begin{align}
\begin{split}
S_{1} =& -27\,{y^2}\,{{{M_a}}^4}\,{{( {y^2} + {{\Delta }^2} - 2\,{M_b} )}^2}
       \,{{{M_b}}^2} + 27\,{y^4}\,{{{M_a}}^3}\,{{{M_b}}^2}\,{{( 2\,{M_a} + 
       {M_b} ) }^2} - 72\,{y^2}\,{{{M_a}}^4}\,{{{M_b}}^2}\,( ( {y^2} + 
       {{\Delta }^2} ) \,{M_a}\\
     & + 2\,{y^2}\,{M_b} - {{{M_b}}^2} ) + 9\,{y^2}\,{{{M_a}}^3}\,( {y^2} + 
       {{\Delta }^2} - 2\,{M_b} ) \,{M_b}\, ( 2\,{M_a} + {M_b} ) \,( ( {y^2} 
       + {{\Delta }^2} ) \,{M_a} + 2\,{y^2}\,{M_b} - {{{M_b}}^2} )\\
     &  - 2\,{{{M_a}}^3}\,{{( ( {y^2} + {{\Delta }^2} ) \,{M_a} + 2\,{y^2}\,
       {M_b} - {{{M_b}}^2} ) }^3} + ({{M_a}^6}\,( ( 27\,{y^2}\,{M_a}\,
       {{( {y^2} + {{\Delta }^2} - 2\,{M_b} ) }^2}\,{{{M_b}}^2}\\
     & - 27\,{y^4}\,{{{M_b}}^2}\,{{( 2\,{M_a} + {M_b} ) }^2} + 
       72\,{y^2}\,{M_a}\,{{{M_b}}^2}\,( ( {y^2} + {{\Delta }^2} ) \,{M_a} 
       + 2\,{y^2}\,{M_b} - {{{M_b}}^2} )\\
     & - 9\,{y^2}\,( {y^2} + {{\Delta }^2} - 2\,{M_b} ) \,{M_b}\,
       ( 2\,{M_a} + {M_b} ) \, ( ( {y^2} + {{\Delta }^2} ) \,{M_a} +
       2\,{y^2}\,{M_b} - {{{M_b}}^2} ) +  2\,( ( {y^2} + {{\Delta }^2} ) 
       \,{M_a}\\
     & + 2\,{y^2}\,{M_b} - {{{M_b}}^2} )^3 )^2 - 4\,( -12\,{y^2}\,
       {M_a}\,{{{M_b}}^2} - 3\,{y^2}\,( {y^2} + {{\Delta }^2} - 2\,{M_b} ) 
       \,{M_b}\, ( 2\,{M_a} + {M_b} )  + ( -( ( {y^2} + {{\Delta }^2} ) 
       \,{M_a} )\\
     & - 2\,{y^2}\,{M_b} + {{{M_b}}^2} ) ^2 ) ^3 ))^{1/2}
\end{split}\\
\begin{split}
S_{2} =& -{S_1^{1/3}} / {2^{1/3}} - ({2^{1/3}} / {S_1^{1/3}}) {{{M_a}}^2}\,
       ( {{\Delta }^4}\,{{{M_a}}^2} + 
       {{\Delta }^2}\,( 2\,{y^2}\,{{{M_a}}^2} - 3\,{y^2}\,{{{M_b}}^2} - 
       2\,{M_a}\,{M_b}\,( {y^2} + {M_b} )  ) \\
     & + {( {{{M_b}}^2} + {y^2}\,( -{M_a} + {M_b} )  ) ^2} )
\end{split}\\
\begin{split}
S_{3} =& ( {y^2} - 2\,{{\Delta }^2} ) \,{{{M_a}}^2} + 
 (3/4)\,{y^2}\,M_b^2 + 
  {M_a}\,{M_b}\,( -{y^2} + 2\,{M_b} )
\end{split}\\
\begin{split}
S_{4} =& y\,( -2\,{M_a} + {M_b} ) \,( 4\,{{\Delta }^2}\,{{{M_a}}^2} + 
    ( {y^2} + 4\,{M_a} ) \,{{{M_b}}^2} ),
\end{split}
\end{align}
where
\begin{equation}
M_a\equiv M+2 M_s-M_{\phi}+M_{\Delta}\quad,\quad
M_b\equiv 3 M_s+2 M_{\Delta}
\end{equation}
Figure \ref{dosgapM6} shows the destructive
effect of the scattering rate produced by fluctuations of $|\Delta|$,
$2\pi\rho_F M_\Delta$, on superconductivity. 
This is very similar to the destruction of superconductivity through the
more standard paramagnetic scattering rate $2\pi\rho_F M_s$. The close
correspondence of these two moments can also be seen by the 
similar way they enter in $M_a$ and $M_b$.
We want to emphasize that the present theory using local averaged one
particle Green's functions represents all basic features of standard type II
superconductivity theory, a fact that has been supported in many details in
previous publications.
With the present extension of this work we want to give the reader the
possibility to see the striking difference between standard paramagnetic 
pairbreaking and related mechanisms in this chapter, and pairbreaking 
induced by the vicinity of spin glass order in the phase diagram, covered
in the following sections.

%% file: scsg0rsb.tex
\section{Competition between superconductivity and spin glass order}
\begin{figure}
\centerline{
\epsfbox{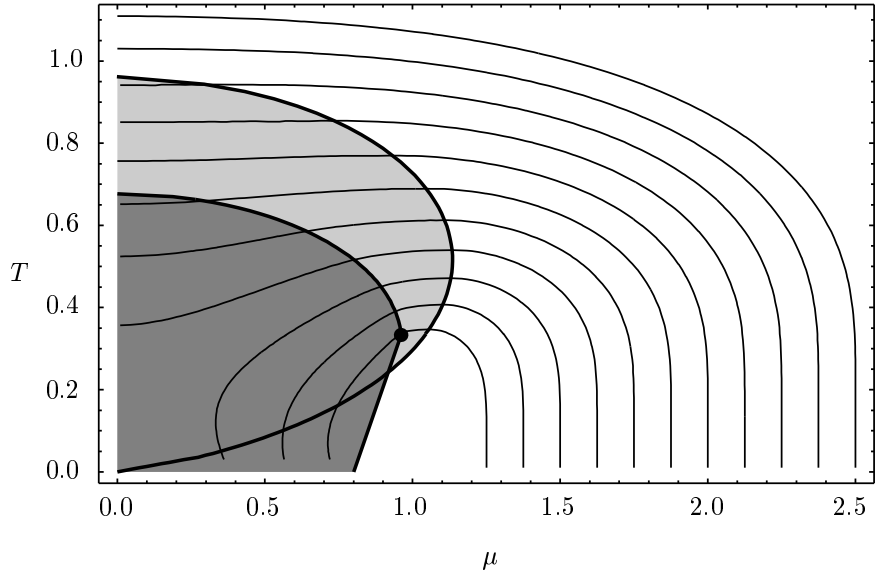}}
\caption{Phase diagram of competing spin glass (SG) and superconducting 
(SC)-transitions (thin lines) for various attractive couplings $v$ 
(given by $v=2J\mu_{e.p.}$ at lines' right endpoint
($T=0,\mu=\mu_{e.p.}$)), at fixed $J=1$, and as a function of
chemical potential $\mu$ (dark grey area: maximum SG-domain for $v=0$).
Thin lines enclose SC-phases, which remove pieces of the
($v=0$) SG-phase or suppress it totally at large enough $v/J$.
The bold line delimiting the light grey area joins
tricritical points ($T_c$-maxima) on SC-critical curves and encloses 1st
order transitions on its left
}
\label{phdiagram0rsb}
\end{figure}

\begin{figure} 
\centerline{
\epsfbox{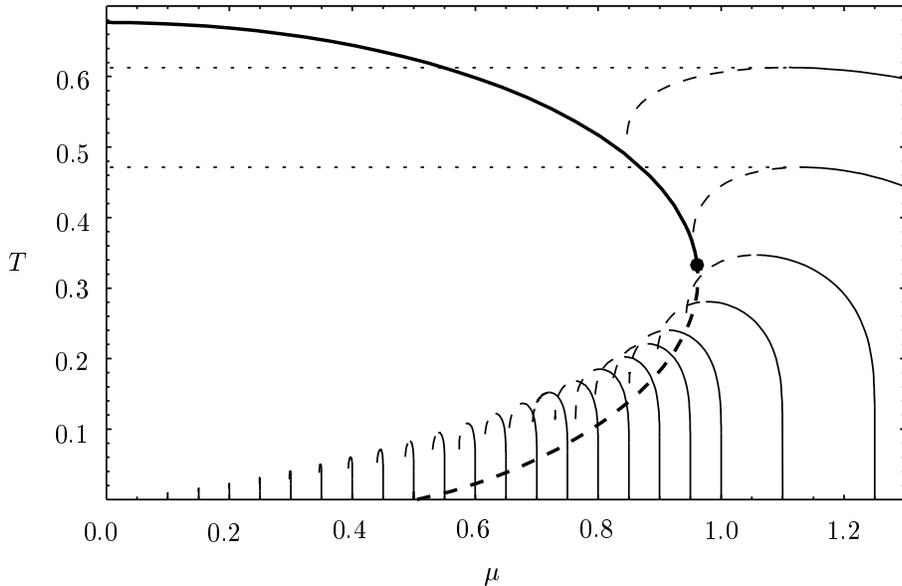}}
\caption{Phase diagram close to the spin glass transition (bold line:
pure SG-transition, bold-dashed line: paramagnetic stability limit)
showing the stability limits of the superconducting (SC) phase (dotted
horizontal lines, given only for $v/J=3.5, 3.0$) and of the non-SC phase
(dashed thin lines) bifurcating at SC-tricritical points.
SC-critical (continued thin) lines are shown down to small $v/J$.}
\label{phdiagram0rsb_complement}
\end{figure}

\newcommand{\qs}{\tilde{q}}
\label{scsg0rsb}
This section is devoted to the solution of the replica symmetric
equations presented for the coexistence problem in section
\ref{scsgbasics}. The phase diagram does not depend strongly
on quantum--dynamical corrections $\chi(\omega)-\chi(0)$, where
$\chi(0)=\beta(\tilde{q}-q)$, which are small for small $\Delta$ and small
ratios $t/J$. The full problem is probably harder to solve than the
infinite--dimensional Hubbard model. Thus approximations are necessary at
present. 
One may view the superconducting order parameter like a generalized
transverse field which induces a quantum-dynamical spin glass.
If one has in mind the analysis of quantum phase transitions and 
the zero temperature limit in particular, then it is
important to study the dynamic mean field theory similarly to the way it
was done in Ref. \onlinecite{fedorovshender} or in Refs. \onlinecite{rsy}
and \onlinecite{rso} for the 
Ginzburg Landau theory.\\
In turn it was mentioned that the existing dynamic theories may not be
able to keep track of nonperturbative phenomena, mentioning the Griffith
singularities as a possible source of concern \cite{rsy}.\\
Thus it is worthwhile to study thoroughly the already difficult
spin/charge--static mean field theory.\\ 
The phase diagram illustrated in Figures \ref{phdiagram0rsb} and
\ref{phdiagram0rsb_complement} is obtained from the free energy of model
${\cal{H}}_{Jv}$. Saddle point solutions $q$, $\tilde{q}$, and $\Delta$
are found from the coupled selfconsistent equations given above.
The physical solutions of these coupled equations are not easy to identify. 
Even outside the SG-phase multiple solutions exist and changeovers of 
stability occur as chemical potential $\mu$, temperature T, or interaction
ratio $v/J$ are varied.
This leads to the complexity of the phase diagram displayed in 
Figures \ref{phdiagram0rsb} and \ref{phdiagram0rsb_complement}. 

Figures \ref{phdiagram0rsb} and \ref{phdiagram0rsb_complement} reveal
unique features of the competition between SG- and SC-order.  For
example, an enhanced fermion concentration $\nu(\mu)$ reduces the
effective spin density at larger $|\mu|$ and is seen to suppress the
spin glass phase stronger than it eliminates the superconducting one
as $\nu(\mu)\rightarrow$ 0 or 2.  Within a large region above the
SG-phase, the SC-critical curves become deformed, as
Fig.\ref{phdiagram0rsb} shows for $v/J<4.5$, due to the increasing
spin glass fluctuations feeded by the random
magnetic interaction. \\
As the critical SC-curve passes through a maximum and starts to
descend with decreasing $\mu$, the SC-transitions change from second
to first order.  For still smaller ratios $v/J$ the SC-line enters the
$v=0$ SG-phase: in this case, magnetic moments freeze first and a
discontinuous SG-SC transition follows at lower temperature (shown for
$v/J=3.25, 3.5, 3.75$).  For still smaller $v/J$ the 1st order SC-line
falls rapidly to zero and the SG-phase prevents superconducting order
up to a critical value $\mu_c$.  The replica symmetric solution leaves
open the possibility of reentrant SC - SG transitions, as one can
observe for $v/J=2.5, 2.75, 3$ in Figure
\ref{phdiagram0rsb}.\\
(remark: if one prefers to include a factor $\frac{1}{2}$ into the
definition of the spin operators (we chose $\hat{\sigma}^z\equiv
n_{\uparrow}-n_{\downarrow}$, $\hbar=1$) a factor of $\frac{1}{4}$
would
rescale the ratios $v/J$ given throughout this paper.)\\

Figure \ref{phdiagram0rsb_complement} complements Figure
\ref{phdiagram0rsb} by adding, within a magnification of the
SG-SC boundary, the stability limits of the superconductor; it
also displays second order SC-transition lines crawling into the phase
separation regime of the Ising spin glass for small $v/J$ and $T$. For
$T\rightarrow0$, the Quantum Parisi Phase must be expected, which means
that the paramagnetic stability limit and the thermodynamic transition
shift to smaller $\mu$. The increase of SG-energy due to replica
symmetry breaking (RPSB), also known from the standard model
\cite{parisi80b}, is evaluated for the fermionic model in the final
chapters. Lacking sufficient information on the full Parisi solution
we employ for the moment numerical solutions of fermionic TAP-equations
\cite{TAP,rehker} to arrive at a straight line estimate of the
discontinuous paramagnet-SG transition curve. 
Discontinuous SC-SG transitions for $\frac{v}{J}<2.5$ must occur in between 
this curve and the SC-SG transition at $\frac{v}{J}=2.5$. 
It is currently not possible to locate better the position of these
discontinuous transitions at low $T$. 
This would require to unite the dynamic mean field theory
with the Parisi solution at infinite RPSB and probably with
Dotsenko--M\'ezard VRSB (despite the discontinuity, one step RPSB may
yield a good approximation but it is not the exact solution at and near
$T=0$). This remains an important research problem for the future. \\
Furthermore, we arrived at the following conclusions:\\
i) There is no coexistence of spin glass - and local pairing 
superconducting order parameter in zero magnetic field. The detailed
analysis of the free energy and of all stability conditions shows only
SC-SG transitions between $\{q\neq0,\Delta=0\}$ and $\{q=0,\Delta\neq0\}$
for $H=0$. We stress that this is concluded from our
$O((t/J)^0)$--calculations covering local pairing superconductivity in a
magnetic band; our experience with metallic spin glasses
\cite{romb,rso} tells us that small $t/J$ will not change the conclusion,
but a large hopping band appearing as a nonmagnetic band,
squeezed in between the two magnetic bands, may allow for the coexisting
SC - SG order parameter. We believe that this interesting and difficult
question is disconnected from the one where coexistence relies on
a special symmetry like d--wave superconductivity. \\ 
ii) The transition between the two phases is always discontinuous
and exists only within a certain range of chemical potentials $\mu$
(or filling $\nu$).\\
iii) For large enough $v/J$, like $v/J>4.13$ at half-filling for
example, the spin glass is prevented by the superconducting transition,
which is continuous for $v/J>4.55$ at half-filling; 
below this value it becomes discontinuous. 
The tricritical line, which separates these domains is included in Figure
\ref{phdiagram0rsb} for several values of $v/J$ and as a function of the
chemical potential.\\
iv) Decreasing the temperature at fixed $\mu<.96$ (see Figure
\ref{phdiagram0rsb}), 
a second transition from SG to superconductivity occurs at
$T_c<T_{f}(v=0)$. 
Figures \ref{phdiagram0rsb} and \ref{phdiagram0rsb_complement} display the
exact numerical results of the replica symmetric theory. 
The spin glass free energy,
increased by only a few percent at higher temperatures, 
rises substantially at smaller temperatures due to RPSB and thus
enlarges the superconducting domain. 
Recalling that reentrant behaviour finally disappears at the spin
glass-ferromagnetic boundary under $\infty$--step RPSB \cite{binderyoung},
the reentrance from superconductor to spin glass and back to the
superconductor seems to disappear already at 1--step RPSB, since the SC-SG
boundary will not become a vertical line in the $(T,\mu)$--plane (unlike
boundaries between SG and (anti)ferromagnetic phases in the
$(T,H)$--plane for magnetic models \cite{binderyoung}). 

\begin{figure}
\centerline{
\epsfbox{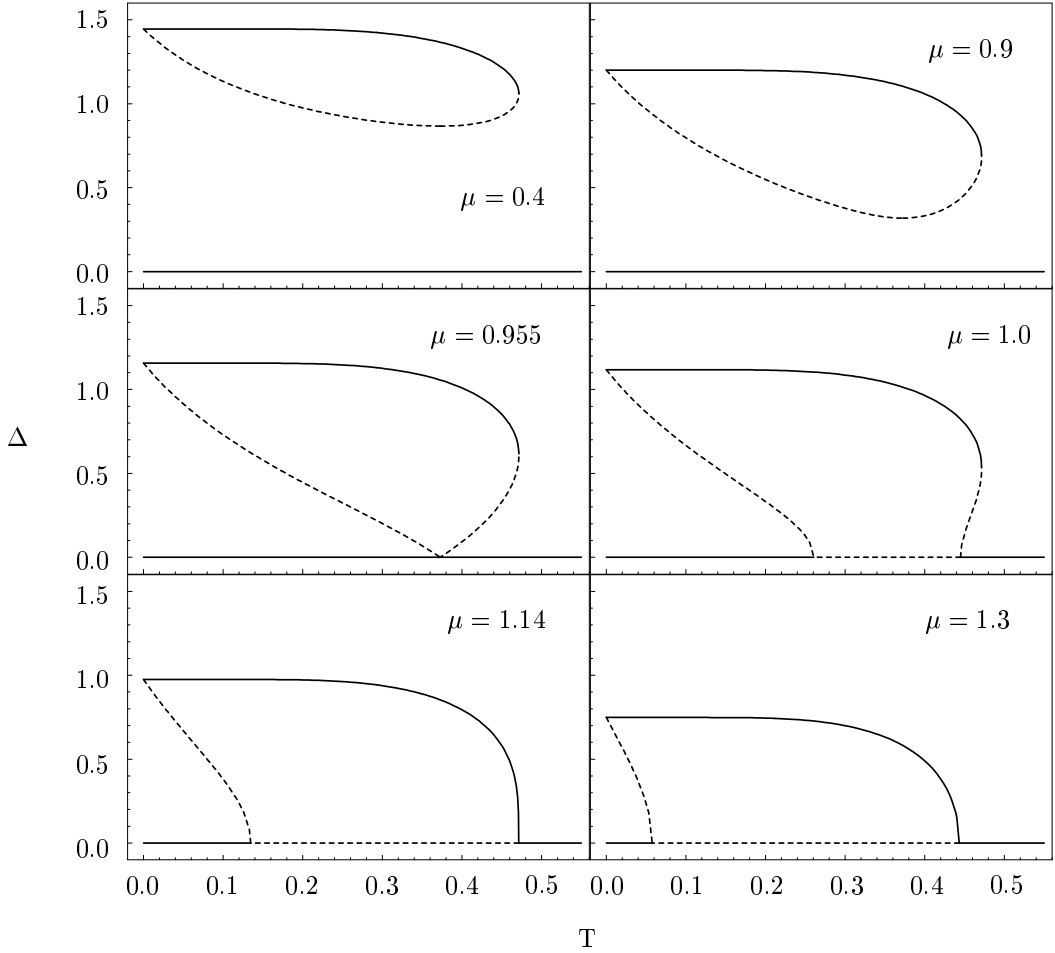}}
\caption{Stable selfconsistent solutions (continuous lines) and
unstable solutions (dashed lines) of the superconducting order parameter
$\Delta$ for $v=3$ (with $J=1$) and $q=0$. Depending on the value of
$\mu$, 4 regions are distinguished: below $\mu=0.955$, a
locally stable solution with $\Delta\neq 0$ emerges at $T_{3} = 0.475$,
and there may or may not be a first order transition, depending on the
free energies of the $\Delta = 0$ and $\Delta > 0$ solutions.
For  $0.955 < \mu < 1.14$, the $\Delta = 0$ solution becomes unstable
in a certain temperature range and a first order superconducting
transition occurs. The second locally stable
$\Delta=0$ solution at low tempatures always has a higher free
energy, there is no second first order solution. For $1.14 < \mu <
\frac{1}{v} = 1.5$ the transition is continuous and both $T_{c}$
and $\Delta(T=0)$ decrease with increasing $\mu$. For higher
$\mu$ the superconducting phase disappears completely.}
\label{DeltaPlots}
\end{figure}

\begin{figure}
\epsfbox{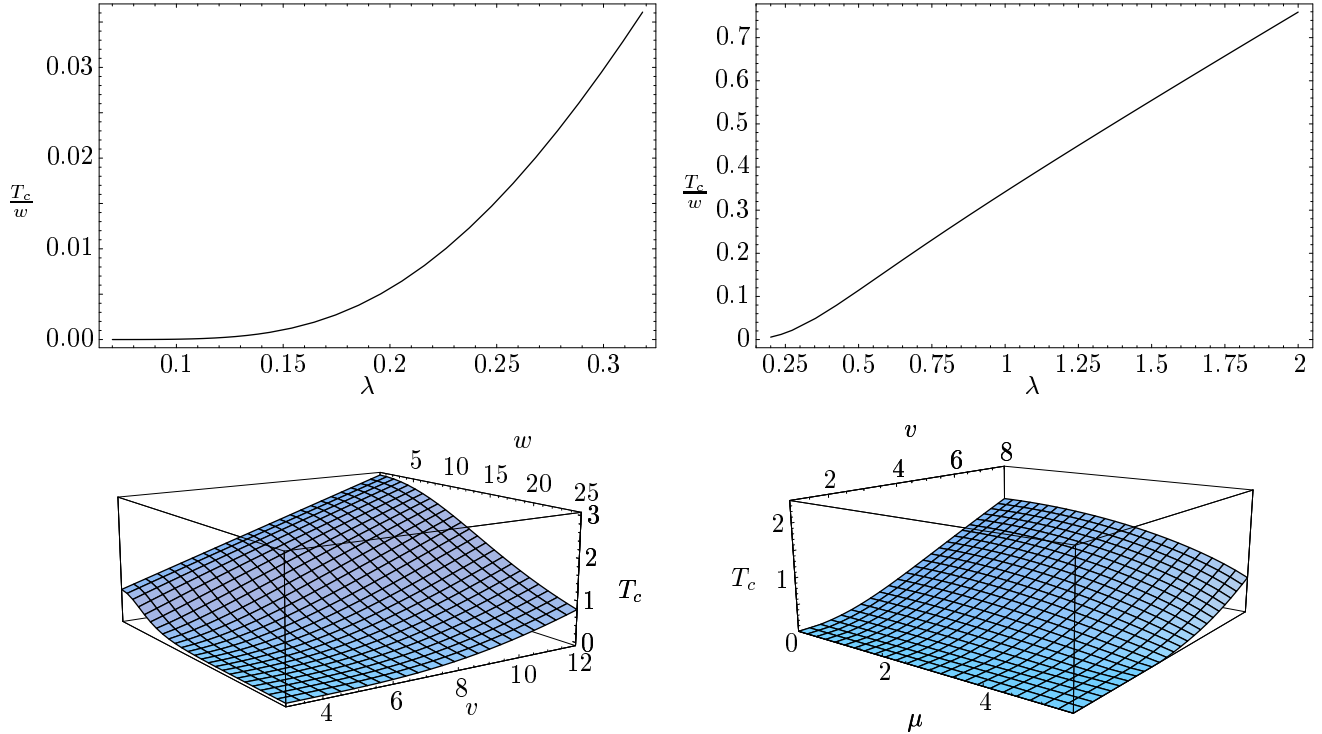}
\caption{upper left: BCS--type exponential decay for small couplings
$\lambda\equiv v\rho_F$; upper right: linear Bose condensation type
$T_c(\lambda)$--regime; lower left:
Complete crossover from Bose condensation type to BCS behaviour
illustrated in 3D plot of $T_c(v)$ as the attractive coupling versus
hopping bandwidth ratio $v/w$ decreases; lower right: BCS--Bose crossover 
also displayed as a function of chemical potential $\mu$} 
\label{bcsbosecrossover}
\end{figure}

Figure \ref{DeltaPlots} shows the effect of random magnetic fluctuations,
described by $\tilde{q}(T,\mu)$, on the superconducting order parameter
for characteristic values of $\mu$ and $v/J=3$, in order to explain the
stability of the different phases and the competition between them.
The absence of coexistent order parameters allows to
set $q=0$. \\
In a magnetic field new aspects arise: the transition temperature of the
superconductor will be reduced and finally vanish for $H> H_{c2}$,
leaving a smeared spin glass transition for sufficiently small
$\mu$. The overlap parameter $q$ is nonzero in a field and
then coexists with $\Delta$, since the field can penetrate the present
type II superconductor. Thus the Almeida Thouless line
can enter the superconductor, infiltrating ergodicity breaking there.\\
We analyze the superconducting phase by means of
the Green's functions: we derive the normal one
${\cal{G}}(\epsilon_n)$ , the anomalous one ${\cal{F}}(\epsilon_n)$,
and relevant particle-hole-$\Pi_{ph}(\omega_m)$ and particle-particle
bubble diagrams $\Pi_{pp}(\omega_m)$ for various limits. Superconductivity
arising in the magnetic band much larger than the hopping band is
described by
\begin{align}
        {\cal{G}}(\epsilon_l) =& \sum_{\zeta = \pm
        \sqrt{|\Delta|^{2}+\mu^{2}}}
        \frac{1}{\qs (\exp(\beta^{2}\qs/2)+\cosh(\beta\zeta))} 
        \frac{\zeta+\mu}{\zeta}\Bigl(
        \frac14 \cosh(\beta\zeta)(i\epsilon_{l} - \zeta)
        \hspace{.1cm}{\cal{U}}\left[1,\frac32,\frac{-(\zeta - i
        \epsilon_{l})^{2}}{2 \qs}\right]\nonumber\\
        &+ \frac18 \sum_{s = \pm 1} (i \epsilon_{l} - \zeta - s\beta\qs)
        e^{\beta^{2}
        \qs/2}\hspace{.1cm}{\cal{U}}\left[1,\frac32,\frac{-(\zeta - i
        \epsilon_{l} + s \beta \qs)^{2}}{2\qs}\right]
        \Bigr)\label{eq:GzeroField} \\
        {\cal{F}}(\epsilon_l) =& \sum_{\zeta = \pm \sqrt{|\Delta|^{2}+
        \mu^{2}}}\frac{\Delta}{\qs (\exp(\beta^{2}\qs/2)+
        \cosh(\beta\zeta))}\frac{1}{\zeta}\Bigl(\frac14
        \cosh(\beta\zeta)(i\epsilon_{l} - \zeta)
        \hspace{.1cm}{\cal{U}}\left[1,\frac32,\frac{-(\zeta - i
        \epsilon_{l})^{2}}{2 \qs}\right]\nonumber\\
        &+ \frac18 \sum_{s = \pm 1} (i \epsilon_{l} - \zeta - s\beta\qs)
        e^{\beta^{2}\qs/2}
        \hspace{.1cm}{\cal{U}}\left[1,\frac32,\frac{-(\zeta -
        i\epsilon_{l} + s \beta \qs)^{2}}{2 \qs}\right]\Bigr)
\label{eq:FzeroField}
\end{align}
Here, $\mathcal{U}$ denotes the hypergeometric $\mathcal{U}$-function
(Kummer function)\cite{AbraStegun}.
Note that the branch cut on the real axis separating the regimes of
retarded and advanced Green's functions is a feature of the Hypergeometric 
$\mathcal{U}$-function.
The SC order parameter obeys
$\Delta=v T\sum_{\epsilon}{\cal{F}}(\epsilon)$.  
\begin{figure}
\centerline{
\epsfbox{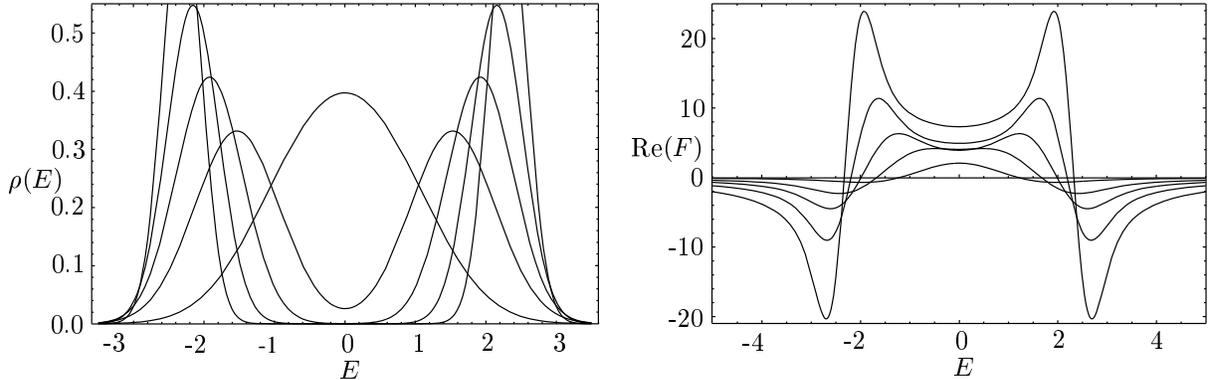}}
\caption{Density of states $\rho(E)$ and ${\rm{Re}}(F(E))$, 
for ($v=5J,\mu=0,T_c\approx 1.11J$) and real energies $E$, 
show crossover from strongly to weakly gapless superconductivity
for decreasing temperatures $T/J=1.1, 1.0, .9, .8$ and $.7$, labeling
$\rho$ from top to bottom at $E=0$ and ${\rm{Re}}(F)$ vice versa.}
\label{Frho}
\end{figure}
The solution for ${\cal{G}}(\epsilon)$ provides (after analytical
continuation) the density of states $\rho(E)$, $E\equiv\epsilon+\mu$. 
A typical result is plotted together with Re$(F)$ 
in Figure \ref{Frho}. Weak fermion hopping--effects and dynamic
corrections from $\Pi_{ph}(\omega)$ and $\chi(\omega)$ are 
negligible here. In principle the frustrated magnetic interaction
interferes by means of a Lorentzian--shaped dynamic susceptibility, which
becomes however $\delta$-like weighted at zero frequency as
$\Delta\rightarrow 0$.
Thus near the phase boundary, where we wish to represent the pairbreaking
effects of the spin glass fluctuations competing with superconducting
order, the given exact solutions of
the $Q$--static model remain sufficiently good approximations. While
the dynamic problem cannot be solved exactly, the given zero-frequency
solution provides a starting point for a new dynamic approximation which
we'll present in a future publication.\\
The plots of Figure \ref{Frho} employ the stable selfconsistent 
solutions for $\tilde{q}(T,\mu)$ and $\Delta(T,\mu)$ inserted in the
(analytic continuation) of solutions given by
Eqs.(\ref{eq:FzeroField}) and (\ref{eq:GzeroField}). 
They demonstrate the crossover from strongly gapless superconductivity
near the spin glass transition to a pronounced pseudogap deeper in the
sc--phase.
Although invisibly small for temperatures lower than roughly 20\% below
$T_c$, the density of states remains nonzero in the pseudogap regime and
vanishes there only at $T=0$. In the strict sense of the word the
superconductor is gapless for all temperatures. But after a transient
shell near the magnetic phase boundary is crossed, the superconducting gap
is almost perfect. One should remark that near the phase boundary, within
the shell where the magnetic susceptibility is not yet strongly depressed,
a piece of the smooth depletion of density of states is also of magnetic
origin. It is not the spin glass gap, but the frustrated interaction tends
to deplete DoS around the Fermi level and at the same time removes the
sharp gap edges which would otherwise appear at all temperatures below the
superconducting transition.\\
The rounding of the superconducting gap edges and its soft progression
below $T_c$ is reminiscent of the magnetic gap found below spin glass
transitions \cite{robrprb} but is of totally different origin.\\ 
Up to now we did not include corrections from the hopping assumed to be
very weak. Fermion hopping of course introduces dynamic effects, and both
the Q--static approximation as well as the quantum--dynamic one had been
worked out before. The smallness of these effects of order $O(t/J)$ does
not change the phase diagram derived for $t/J=0$. One may calculate
peturbatively corrections as done in Ref. \onlinecite{romb}. 
We refrained from doing
this because it is only important for the quantum phase transition, which
requires however the full Parisi solution and the analysis of vector
replica symmetry breaking as discussed before.\\
As one starts to look deeper into the superconducting phase, 
the magnetic band narrows however
and eventually the fermion hopping (bandwidth) starts to dominate.
Then, the normal and the anomalous Green's function cross over into
the hopping band solution given by Eqs.(\ref{hopping_G}) and
(\ref{hopping_F}), respectively.
The transition temperature derived from these solutions illustrates in
Figure \ref{bcsbosecrossover} the crossover from linear Bose
condensation $T_c(v)$-dependence to exponential BCS-type
superconductivity.
The Bose-BCS crossover can be followed through the whole parameter range 
due to the local property of the one-particle Green's functions, a feature
appreciated as well in the $d=\infty$ method for clean systems
\cite{vollhardt,GeKr}. 
The suppression of one particle phase coherence, due
to infinite dimensions in clean systems or by symmetry
requirement in the quenched average used here, results in type II
superconductivity with the two particle coherence length replacing the
standard one in the Ginzburg Landau theory (many details including 
paramagnetic pairbreaking were published in Ref. \onlinecite{oppermann}).
The puzzling coexistence phenomena in zero field, manifested by the
absence of stable $(q\neq0,\Delta\neq0)$-solutions and by nonexisting
common onset of magnetic and superconducting order on one hand and 
coexistence within phase separation regimes on the other, is further
elucidated by the bubble diagram $\Pi_{pp}(T,\omega)$. 
The critical condition $1=v\Pi_{pp}(T_c,\omega=0)$
can only be satisfied for $v>v_{min}$, with $v_{min}=4.14J$ at half
filling. One can solve for $T_c=T_f$, finding $v=4.186J$ at half filling 
for example, but the second solution at $T_c'>T_c$ for this coupling
$v$ turns out to be the stable one. The thermodynamic transition occurs at
a still higher temperature $T_{c1}>T_c'$. Thus the calculation based on
the two-particle Green's functions confirms the absence of simultaneous
and continuous onset of spin glass and superconducting order.\\  
For the sake of transparency we discussed only fully frustrated magnetic
interactions and zero field phenomena. Subsequent antiferromagnetic-spin
glass-superconductor transitions, as $\mu$ increases are not
simply obtained once $J_{ij}$ contains an antiferromagnetic part. 
A model extension by the Hubbard interaction can change this;
allowance for different symmetries of the order parameter, and the
dynamic quantum Parisi phase
are further examples for future research on links between
antiferromagnetism, spin glass order, and superconductivity.\\ 
The model we analyzed here proved to have a characteristic crossover between 
strongly and weakly gapless superconductivity, due to correlations induced 
by the frustrated magnetic interaction and depending on the vicinity of a
spin glass phase. 


%% file: scsg1rsb.tex
\section{Effects of replica permutation symmetry breaking on the spin
glass - superconductor phase boundary}
\label{scsg1rsb}
\begin{figure}[htbp]
\centerline{
  \epsfbox{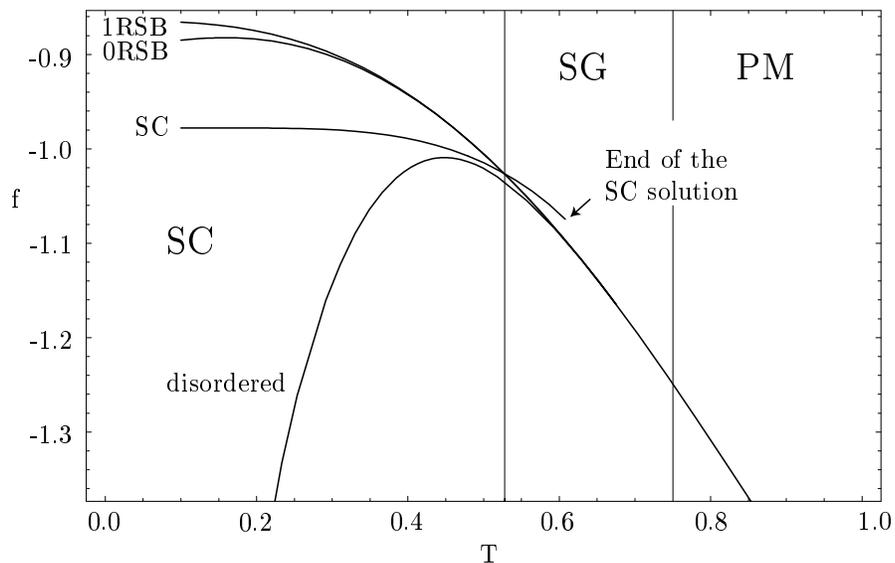}}
  \caption{Free energies $f$ of the disordered and spin glass solutions,
    together with the superconducting solution for $v=3.5$. It is easy to
    infer the SC--SG first order transition.
    At the second order transition SG--PM,
    the free energies of the spin--glass and the disordered solutions merge.
    Note that the disordered solution is unstable with respect to $q$ below
    this transition, therefore it is not the physical one although it has the
    lowest free energy. Also clearly visible is the difference between the
    1--step RSB solution and the replica--symmetric one. It becomes
    important at low temperatures.}
  \label{fig:fplot01}
\end{figure}

\begin{figure}[htbp]
\centerline{
  \epsfbox{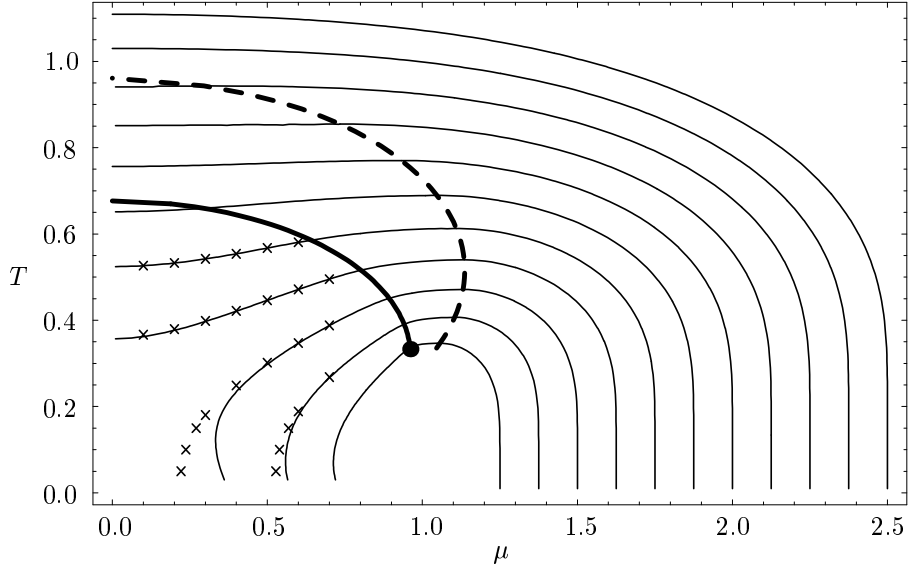}}
  \caption{Phase diagram including 1--step RPSB corrections to the phase
  diagram of Figure \ref{phdiagram0rsb}. 
  Crosses indicate points of the SG--SC boundary calculated in 1--step
  RPSB. 
  The deviations from the RS--calculations are most significant for low
  temperatures and suppress reentrant transitions.}
\label{fig:diagram1RSB}
\end{figure}

Despite the absence of coexisting order parameters, the phase boundary
depends on replica symmetry breaking due to the change of the spin glass
free energy. This change becomes important at low temperatures. There, the
increase in free energy entrained by RPSB becomes large enough to suppress 
reentrant behaviour.  
In Figure \ref{fig:fplot01} the free energies are shown for $\mu=0.1$ and
for a
moderately large attractive coupling $v=3.5$: the left upper corner shows
the increasing deviation and enhancement of the free energy with 1RPSB.
The clearly visible crossing between the SC-curve and the spin glass curve
indicates the thermodynamic discontinuous phase transition from spin
glass in the intermediate temperature while the SG - PM transition is
shown where the corresponding (and almost indistinguishable on the given
scale) free energies become equal. 
Another example is given in the following Figure \ref{fig:fplot03} for
$\mu=0.3$.
The consequences of 1RPSB--corrections on the phase diagram as shown in
these figures are quite clear. As the free energies exceed
substantially the replica-symmetric ones in the low temperature regime,
where reentrance was observable in 0RPSB, the 1RSB--corrections strongly
suppress reentrance. A first indication of this is shown in
Figure \ref{fig:diagram1RSB}.

\begin{figure}
\centerline{
  \epsfbox{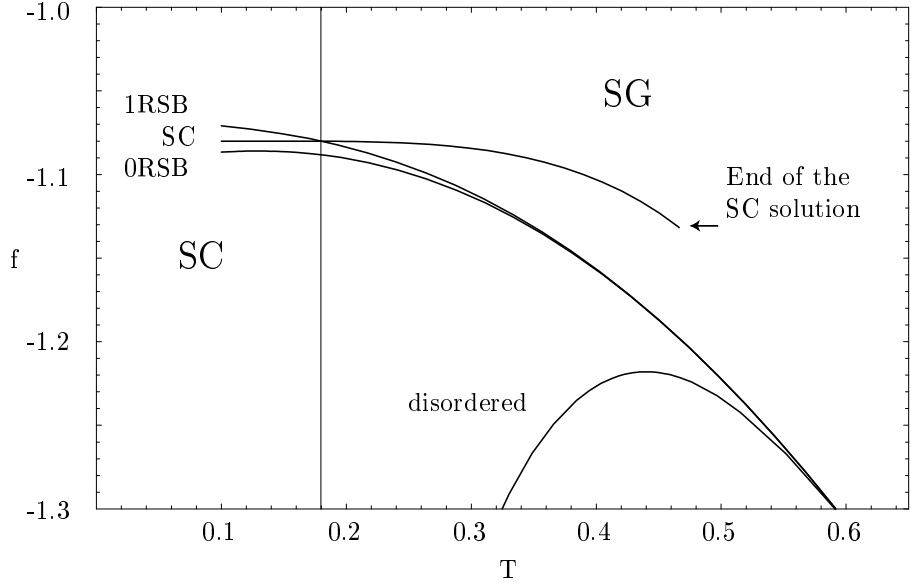}}
  \caption{Free energies $f$ of the disordered, spin glass, and
    superconducting solutions as in Fig. \ref{fig:fplot01}, for
    $\mu=0.3$ and $v=3.0$. Here one sees clearly how the SG-SC first order
    transition is affected by RPSB. Also, reentrance from SC to SG is
    suppressed with increasing RPSB, because the maximum of the SG
    solution
    becomes less pronounced and shifts to lower temperatures.
        For infinite RPSB we expect it to be at $T=0$.}
  \label{fig:fplot03}
\end{figure}

\begin{figure}
\centerline{
\epsfbox{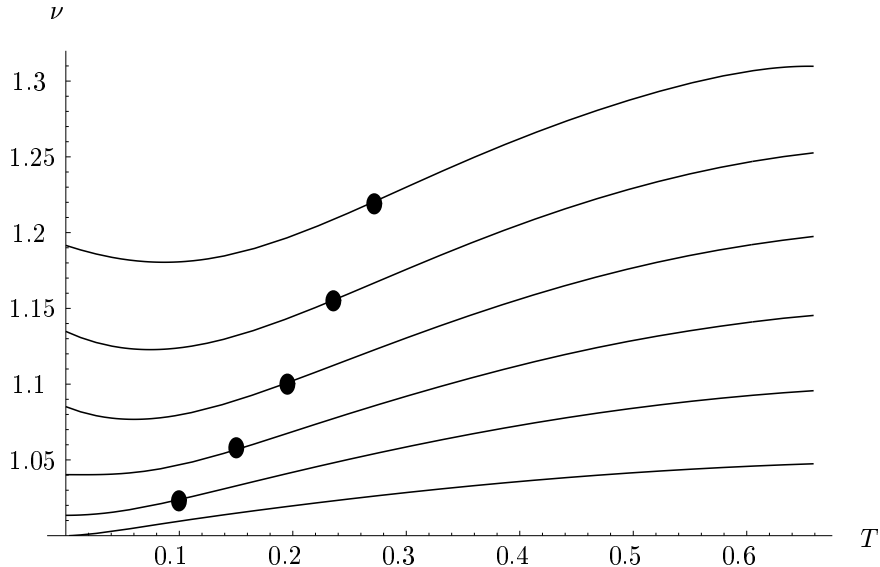}}
\caption{The fermion concentration $\nu(\mu)$ in 1--step RPSB as a
function of temperature, shown for chemical potentials $\mu=.6,
.5,...,.1$ from top to bottom. Dots indicate the location of the
random field crossover line.}
\label{filling}
\end{figure}

The numerical effort in calculating 1RPSB results at very low temperatures
is high and we shall report elsewhere a more detailed analysis employing
a more dense set of chemical potentials.
\\
For a certain set of chemical potentials we also calculated the fermion 
filling factor as a function of temperature as shown in
Figure \ref{filling} for 1--step RPSB. The turnaround at low temperatures
into a filling which decreases as the temperature increases from zero
may belong to the unstable regime below the AT--instability--line, while
the dots are 1step--RPSB results taken from the crossover line
$\partial^2 F/\partial\tilde{q}^2=0$. The curve shown for $\mu=.1$
and any other curve with $\mu<.119$ lies entirely in the stable
regime at 1--step RPSB, but this changes with the order $k$ of RPSB,
since the gap decreases to zero \cite{eplrobr} as
$k\rightarrow\infty$.

%% file: pam.tex
\section{Discussion, outlook, open problems}
\label{pam}
\subsection{Expectations on lower dimensional corrections: which mean
field predictions are robust?}
Dimensional dependences can lead to inapplicability of mean field theories
to lower-dimensional systems in a way similar to the failure of numerical
results on too small systems.\\ 
While correlation functions explore phase transitions with high
sensitivity and thus depend strongly on dimensions, particularly as those
drop below their upper or lower critical values, there are less sensitive
quantities like energies depending much less on these complications. Only
their derivatives are more or less sensitive. 
Also gaps of excitation spectra which are related to the competition of
different scales, can be rather robust.
\subsection{A comparison with the gap structure of the twodimensional
periodic Anderson model}
The spin glass generated charge gap of the fermionic Ising spin glass has
been shown to have important consequences and we considered a gap digged
out by coexisting magnetic and superconducting gap (despite noncoexistence
in zero magnetic field). We find remarkable the fact that the gap
structure obtained for either a fictitious or a finite-field-driven
coexistence of spin glass order and superconductivity on one hand is
comparable with the gap structure of the twodimensional periodic Anderson
model \cite{vekicscalapino}. The latter was derived by the Quantum Monte
Carlo method for a twodimensional model. Clearly the mixed valence
coupling corresponds to our superconducting order parameter in the role as
a gap generator.
One may pick two arbitrary values of $q$ and $\Delta$, not necessarily
selfconsistent, to match the gap structure seen by V\'ekic et al
\cite{vekicscalapino} for the PAM model.
We stress once more that this is not derived selfconsistently and
its interest lies in the gap structure enabled by the functional
dependence of the present model, the possibility to match a twodimensional
systems gap structure by the present mean field theory, and the
possibility that a magnetic field may realize the structure.\\
So we observe several models with rather high analogies:
The periodic Anderson model, where the competition
between the magnetic moment quenching Kondo effect and the RKKY
interaction is important, its disordered version, which also falls into 
the class of randomly interacting models, the present SG - SC
competition in a magnetic band model, and its smeared three band version.
The relation between these models deserves further analysis.
We shall invent
for this purpose a technique which allows to treat dynamic interaction
effects properly.
\subsection{Hopping band: the effective three-band Ising spin glass}
We considered the limit of a magnetic band large compared to the fermion
hopping band. This parameter range refers to transitions between a very
bad conductor and superconductivity which stems from pairing in the
magnetic band(s). We discussed in detail the frustrated magnetic
interaction being at the origin of these magnetic bands, and their decay
deep in the superconducting phase as well. Since the interesting parameter
range for competition between spin glass and superconductivity requires
the corresponding interactions $J$ and $v$ to be roughly of the same
order, and since we deliberately restricted the analysis to very small
$t/J$, the Bose condensation type of superconductivity was considered.
Competition between spin glass and BCS--type superconductivity - within
the same theory - can be investigated under the condition of a large
hopping band. This refers to a nonmagnetic third band occupying
essentially the space evacuated by spin glass order between
the two magnetic bands; strong dynamic effects emerge and many important
features may become different from the one described here: 
Theories for effects of comparable fermion hopping 
in metallic spin glasses exist \cite{rso,romb}, but the
coexistence with BCS-type superconductivity is an open question,
since pairing may occur in the nonmagnetic band. The strong overlap with
the magnetic bands sustaining spin glass order does however not seem to
allow for an intuitive prediction on the fate of order parameter
coexistence.
\section{Acknowledgements}
This research was supported by the Deutsche Forschungsgemeinschaft under
research project Op28/5--1 and by the SFB410. One of us (H.F.) also wishes
to acknowledge support by the Villigst foundation.
We are grateful for discussions with J.A. Mydosh, B. Rosenow, and F.
Steglich.